%% file: after-review.tex
\documentclass{article}
\usepackage{graphicx} 
\usepackage{amsmath} 
\usepackage{amssymb} 
\usepackage{color}

\newcommand{\bbibitem}{\bibitem}

\newcommand{\llabel}[1]{{\label{#1}}}

\newcommand{\ffoot}[1]{}

\renewcommand{\r}[1]{(\ref{#1})}
\newcommand{\bi}{\begin{itemize}}
\newcommand{\ei}{\end{itemize}}
\newcommand{\bd}{\begin{description}}
\newcommand{\ed}{\end{description}}
\newcommand{\be}{\begin{enumerate}}
\newcommand{\ee}{\end{enumerate}}

\renewcommand{\i}{\item}

\newcommand{\bqn}{\begin{eqnarray}}
\newcommand{\eqn}{\end{eqnarray}}
\newcommand{\eqnn}{\nonumber\end{eqnarray}}
\newcommand{\eqnl}[1]{\llabel{#1}\end{eqnarray}}

\newcommand{\nn}{\nonumber}
\newcommand{\noi}{\noindent}
\newcommand{\ba}[1]{\begin{array}{#1}}
\newcommand{\ea}{\end{array}}

\newcommand{\R}{\mathbb{R}}
\newcommand{\C}{\mathbb{C}}

\newcommand{\fine}{\end{document}}

\renewcommand{\frak}{\mathfrak}

\def \trait (#1) (#2) (#3){\vrule width #1pt height #2pt depth #3pt}
\def \qed{\hfill
        \trait (0.1) (6) (0)
        \trait (6) (0.1) (0)
        \kern-6pt   
        \trait (6) (6) (-5.9)
        \trait (0.1) (6) (0)
\medskip}
\def \qedmio{\hfill
             \trait (8) (8) (-0.1)
             \medskip}
\def \quadp{{\Huge $\qedmio$}}

\newtheorem{ml}{\bf Lemma}
\newtheorem{Theorem}{\bf Theorem}

\newtheorem{mrem}{\bf \underline{{\sl Remark}}}
\newtheorem{mcc}{\bf Corollary}
\newtheorem{Definition}{\bf Definition}
\newtheorem{mpr}{\bf Proposition}
\newtheorem{mproperty}{\bf Property}

\newcommand{\bt}{\begin{Theorem}}
\newcommand{\et}{\end{Theorem}}
\newcommand{\bl}{\begin{ml}}
\newcommand{\el}{\end{ml}}
\newcommand{\bp}{\begin{mpr}}
\newcommand{\ep}{\end{mpr}}
\newcommand{\bc}{\begin{mcc}}

\newcommand{\bproperty}{\begin{mproperty}}
\newcommand{\eproperty}{\end{mproperty}}

\newcommand{\ec}{\end{mcc}}
\newcommand{\bdeff}{\begin{Definition}}
\newcommand{\edeff}{\end{Definition}}
\newcommand{\brem}{\begin{mrem}\rm}
\newcommand{\erem}{\end{mrem}}

\newcommand{\proof}{{\bf Proof. }}
\newcommand{\ppotR}[3]
{

\begin{figure}\begin{center}
~\includegraphics[width=#3truecm]{./#1.eps}\\
\caption{#2}
\llabel{#1}
\end{center}
\end{figure}
\noindent$\!\!$}
       
\newcommand{\lam}{\lambda}
\newcommand{\g}{\gamma}
\newcommand{\al}{\alpha}
\newcommand{\eps}{\varepsilon}

\newcommand{\con}{{\cal C}}

\newcommand{\F}{{\cal F}}

\newcommand{\sceq}{Schr\"{o}dinger equation\ }  
  
\newcommand{\neigh}{neighborhood }

\newcommand{\e}{\mbox{e}}

\newcommand{\parti}[1]{\frac{\partial}{\partial #1}}

\newcommand{\HHH}{{\cal H}\!\!\!\!{\cal H}}

\newcommand{\und}{\underline}
\newcommand{\da}{\Delta_A^{-1}(0)}
\newcommand{\dbu}{\Delta_{B_1}^{-1}(0)}
\newcommand{\dbd}{\Delta_{B_2}^{-1}(0)}
\newcommand{\dbi}{\Delta_{B_i}^{-1}(0)}

\newcommand{\Da}{\Delta_A}
\newcommand{\Dbu}{\Delta_{B_1}}
\newcommand{\Dbd}{\Delta_{B_2}}

\newcommand{\Sur}{  {\cal S}   }
\newcommand{\Tur}{  {\cal T}   }

\newcommand{\So}{  {\cal S}  }

\begin{document} 

\begin{center} \noindent
{\LARGE{\sl{\bf Nonisotropic 3-level Quantum Systems: Complete 
Solutions for Minimum Time and Minimum Energy
}}}

\vskip 1cm
Ugo Boscain,\\
{\footnotesize SISSA, via Beirut 2-4 34014 Trieste, Italy}\\
Thomas Chambrion\\
{\footnotesize
SYSTeMS Research Group, University of Ghent
    Technologiepark - Zwijnaarde 9, 9052 Zwijnaarde, Belgium,}\\
Gr\'egoire Charlot\\
{\footnotesize
 ACSIOM, I3M, CC51,
Universit\'e©  Montpellier II,
Place Eug\`e¨ne  Bataillon,
34095 Montpellier Cedex 5, France}\\
e-mails {\tt boscain@sissa.it, 
Thomas.Chambrion@ugent.be, 
charlot@math.univ-montp2.fr}
\end{center}

\vspace{.5cm} \noindent \rm

\begin{quotation}
\noindent  {\bf Abstract}
We apply techniques of subriemannian geometry on Lie groups 
and of optimal synthesis on 2-D manifolds 
to the population transfer problem in a three-level 
quantum system driven by two laser pulses, of arbitrary 
shape and
frequency. In the rotating wave 
approximation, we consider a nonisotropic  model, i.e., a model in 
which the two coupling constants of the lasers are different. The aim
is to induce transitions from the first to the third level, minimizing 
{\bf 1)} the time of the transition (with bounded laser amplitudes), 
{\bf 2)} the energy transferred by lasers to the system (with fixed final 
time). 
After reducing the problem to real variables,
for the purpose {\bf 1)} we develop a theory of time optimal syntheses for
distributional problem on 2-D manifolds, while for the purpose 
{\bf 2)} we 
use techniques of subriemannian geometry on 3-D Lie groups. The complete 
optimal syntheses are computed.
\end{quotation}


\section{Introduction}
\subsection{Statement of the Problem}
\llabel{s-intro}
 The problem of designing an efficient transfer of population between
different
atomic or molecular levels is crucial in many atomic-physics
projects \cite{bts,car1,car2,shorebook}.
Often excitation or ionization
is
accomplished by using
a sequence of laser pulses to drive transitions from each state to the 
next
state.  The transfer should be as efficient as possible  in order to
minimize the effects of relaxation or decoherence that are always present.
In the recent past years, people started to approach the design of laser
pulses by using Geometric Control Techniques (see for instance
\cite{rabitz,daless,brokko,rama}).
Finite dimensional closed quantum systems are in fact left (or 
right) invariant control systems on $SU(n)$, or on the 
corresponding 
Hilbert sphere $S^{2n-1}\subset\C^n$, where $n$ is the number of atomic or
molecular  levels. For these kinds of systems very powerful techniques
were
developed  both for what concerns controllability
\cite{G-a,G-gb,G-jk,yuri}
and
optimal control \cite{agra-book,libro,jurd-book}.

The most important and powerful tool for the study of optimal trajectories
is the well known Pontryagin Maximum Principle (in the following PMP, see
for instance
\cite{agra-book,jurd-book,pontlibro}). It is a first order
necessary condition for optimality and generalizes the Weierstra\ss \
conditions of Calculus of Variations to problems with non-holonomic
constraints. For each optimal trajectory, the PMP
provides a lift to the cotangent bundle that is a solution to a suitable
pseudo--Hamiltonian system.
Anyway, giving a complete solution to an  optimization problem (that for
us means to give an
\underline{optimal} \underline{synthesis}, 
see for instance
\cite{boltianski,libro,brun1,piccoli-sussmann}) remains
 extremely difficult for several reasons. First, one is faced with the
problem of integrating a Hamiltonian system (that generically is
not  integrable excepted for very special costs). Second, one should manage
with ``non Hamiltonian solutions'' of the PMP, the so called
\underline{abnormal
extremals}. Finally,  even if one is able to find all the solutions of the
PMP, it remains the problem of \underline{selecting} among them the
\underline{optimal trajectories}.
For these reasons, usually, one can hope to find  a complete solution to
an optimal control problem  in low dimension only 
(\cite{mario,libro,quattro,sch1}).

\bigskip

In this paper we apply techniques of subriemannian geometry on Lie
groups and of optimal synthesis on 2-D manifolds to the 
population 
transfer problem in a three-level 
quantum
system driven by two external fields (in the rotating wave approximation)
of arbitrary shape and frequency.
The dynamics is governed by the time dependent Schr\"odinger equation (in
a system of units such that $\hbar=1$)
\bqn
i\frac{d\psi(t)}{dt}=H\psi(t),
\eqnl{se}
where $\psi(.)=(\psi_1(.),\psi_2(.),\psi_3(.))^T:[0,T]\to\C^3$, 
$\sum_{j=1}^3|\psi_j(t)|^2=1$ (i.e., $\psi(t)$ belongs to the sphere 
$S^5\subset \C^3$), and
\bqn
H=\left(\ba{ccc} E_1&\mu_1 \F_1&0\\ \mu_1\F_1^\ast 
&E_2&\mu_2\F_2\\
0&\mu_2\F_2^\ast&E_3 \ea\right),
\eqnl{hg1}
where $E_1$, $E_2$ and $E_3$ are real numbers representing the \underline{energy levels}.
Here $(^\ast)$ denotes the complex conjugation involution.
The \underline{controls} $\F_1(.),\F_2(.)$,  that we assume  to be 
measurable functions,
different
from zero only in a fixed interval, are the \und{external} \und{pulsed 
field}, 
while 
$\mu_{j}>0,$ $(j=1,2)$ are the couplings
(intrinsic to the quantum system) that we have restricted to
couple only levels $j$ and $j+1$ by pairs.

\bigskip
Using standard arguments 
of controllability on compact Lie
groups and corresponding homogeneous spaces, one
gets the following (see the recent survey \cite{yuri} or the papers
\cite{G-a,q1,q2,G-gb,G-jk}):
\bp
Assume that  there exists constants $M_1,M_2>0$ such that $|u_1(t)|\leq 
M_1$ and  $|u_2(t)|\leq M_2$ for a.e. $t>0$.
Then  there exists a time $\tau(M_1,M_2)$ such that
the control system \r{se}, \r{hg1} is completely
controllable in time $\tau(M_1,M_2)$.
\ep

The  model \r{se}, \r{hg1} belongs to a class of systems on which it 
is 
possible to eliminate the so called drift term (i.e., the 
term $diag(E_1,E_2,E_3)$) by a unitary change of coordinates and a change 
of controls (see Section \ref{s-rrv}).

The aim is to induce complete population transfer from the 
\underline{state one} ($|\psi_1|^2=1$), called \underline{source}, to 
the \underline{state 
three} ($|\psi_3|^2=1$), called \underline{target}, 
minimizing the criteria described in the following.\\\\
{\bf Energy in fixed time}
\bqn
\int_0^T\left(|\F_1(t)|^2+|\F_2(t)|^2 \right)~dt,
\eqnl{nonis-en}
This cost is the energy of the laser pulses.
After elimination of the drift and reduction to a real problem (see below),  
the problem of minimizing this cost becomes a singular-Riemannian
problem (or a subriemannian problem when lifted on the group $SO(3)$, 
see Section \ref{s-rrv}) and was studied in
\cite{q2}, in the 
``nongeneric'' case in which $\mu_1=\mu_2$. In the following we call the 
problem \r{se}, \r{hg1}, \r{nonis-en}  \underline{isotropic}  if 
$\mu_1=\mu_2$ otherwise we call the problem
\underline{nonisotropic}.
For the cost \r{nonis-en}, to guarantee the existence of minimizers, the 
final time 
$T$ should be fixed, and such  minimizers are parameterized 
with constant
velocity ($|\F_1|^2+|\F_2|^2=const$). Moreover  
\underline{the controls are not assumed
a priori bounded}. Anyway  if the final
time $T$ is fixed in such a way the minimizer is parameterized by
arclength ($|\F_1|^2+|\F_2|^2=1$), then minimizing the cost
\r{nonis-en} is equivalent to minimize the time, with moduli of controls 
constrained in the
closed set
\bqn
|\F_1|^2+|\F_2|^2\leq1. \label{titi}
\eqn
For this cost we can normalize
$\mu_1=\mu_2=1$ in the isotropic case, while 
$\mu_1=1$, $\mu_2=\al>0$ for the nonisotropic case.
Notice that this change of notation 
modifies the costs 
only by an multiplicative constant.\\\\
{\bf Time with bounded controls}\\
Another interesting cost is the time of transfer under the conditions
\bqn
|\F_1(t)|\leq\nu_1\mbox{ and } |\F_2(t)|\leq\nu_2.
\eqnl{nonis-t}
In this case we call the problem \underline{isotropic} if 
$\mu_1\nu_1=\mu_2\nu_2$. Otherwise we call the problem 
\underline{nonisotropic}.
Let us notice that 
if we do not require the constraints \r{nonis-t},
then there is no 
minimizer.\\
For this cost we can normalize
$\mu_1=\nu_1=\mu_2=\nu_2=1$
in the isotropic case, while 
$\mu_1=\nu_1=1$, $\mu_2=\al>0$, $\nu_2=1$
for the nonisotropic case.\\\\
After these normalizations the Hamiltonian and the costs read
\bqn
&&H=\left(\ba{ccc} E_1&\F_1&0\\ \F_1^\ast 
&E_2&\al\F_2\\
0&\al\F_2^\ast&E_3 \ea\right),\llabel{h-al}\\
&&\mbox{energy in fixed 
time:}~~~
\int_0^T\left(|\F_1(t)|^2+|\F_2(t)|^2 \right)~dt,
\llabel{energy-al}\\
&&\mbox{time with bounded controls:}~~~ T=\int_0^T 1~ dt, \mbox{ with   
$|\F_1(t)|\leq1$  and  
$|\F_2(t)|\leq1$}.\llabel{time-al}
\eqn
In the following we call the parameter $\al$ the \und{nonisotropy 
factor}.
\brem
\llabel{r-lift}
The problem of inducing a transition from 
the first to the third
eigenstate, can be formulated, as
usual, at the level of the wave function $\psi(t)$, but also at  the level 
of the time evolution operator (the resolvent),
 denoted here by $g(t)$. We have $\psi(t)=g(t)\psi(0),$ $g(t)\in 
U(3),$ $g(0)=id.$ For $g(t)$ the \sceq \r{se} reads 
$\dot g(t)=-iHg(t)$.
In the following we call the optimal control problem for $\psi(t)$
and for $g(t)$ respectively  the \underline{problem downstairs} and 
the
\und{problem upstairs}. In this kind of problems one can take advantages 
of 
working both 
upstairs and downstairs depending on the specific task. This approach 
happened to be successful in some other problems of optimal control 
on Lie groups, see for instance  
\cite{dubin}.
Notice that the evolution of the trace part is decoupled from the rest. It 
follows that we can always assume $E_1+E_2+E_3=0$ and $g(t)\in SU(3)$.
\erem 
This paper is the continuation of a series of papers on 
optimal control of finite dimensional quantum systems 
\cite{q1,q2,q3}.
In \cite{q1}, the problem of minimizing the energy of the lasers pulses
was studied for a two-level system  and for an  isotropic three-level 
system. The two-level system (that is a problem on $SU(2)$) was 
completely solved (in this case minimizing energy is equivalent to 
minimize time with bounded controls), while the three-level problem (that 
upstairs is a problem on $SU(3)$) was solved assuming 
resonance as hypothesis (i.e., controls oscillating with a frequency equal 
to the 
difference of the energy levels that the  laser is coupling, see 
formula \r{res} below). 
Thanks to this assumption, the problem (downstairs) could be reduced to a 
two-dimensional 
problem and solved completely.
In \cite{q2} the problem of minimizing the isotropic energy in 
the three-level problem was 
treated without  assuming resonance as hypothesis. In that case, even if 
the optimal  control problem lives in a space of big dimension 
($dim(SU(3))=8$), it was possible to get  explicit expressions of optimal 
controls and trajectories thanks to the special structure of the 
Lie algebra of the problem and to the fact that the cost is built with the 
Killing form, 
that renders the Hamiltonian system associated to the PMP Liouville 
integrable. In that paper resonance was obtained as consequence of the 
minimization process and explicit expressions of amplitudes  were 
provided.
In \cite{q3}, the possibility of restricting to resonant
controls was generalized to $n$-level systems and to more general costs. 
More precisely consider a $n$-level system of the kind  
\bqn
\left\{
\ba{l}
i\mbox{{\Large $\frac{d\psi(t)}{dt}$}}=(D+V(t))\psi\\\\
\psi(.):=(\psi_1(.),...,\psi_n(.))^T:[0,T]\to\C^n,~~~
\sum_i|\psi_i|^2=1,
\ea\right.
\eqnn
where $D=diag(E_1,...,E_n)$ and $V(t)$ is an Hermitian matrix
$(V(t)_{j,k}=V(t)_{k,j}^\ast)$, whose
elements are either identically zero
or controls.
(i.e., $V_{i,j}\equiv 0$ or $V_{i,j}=\mu_{i,j}F_{i,j}$, 
for $i<j$, the term $F_{i,j}$ being the external pulsed field coupling 
level $i$ and level $j$). Then for a convex cost depending 
only on the moduli of controls (i.e., amplitudes of the lasers), like 
those described above,  it was proved that there  always exists a 
minimizer in resonance that  connects a source and a target
defined by conditions on the moduli of the components of the wave
function (e.g. two eigenstates): 
$\F_{j,k}(t)=u_{j,k}(t)e^{i[(E_{j}-E_{k})t +\xi_{j,k}]},~~i<j=1,...,n,$
where $\xi_{j,k}\in[-\pi,\pi]$ are some phases and 
$u_{j,k}(.):[0,T]\to\R$ 
are the amplitudes of the lasers that should be 
determined.
This result permits to reduce the problem to real variables (i.e., from 
$SU(n)$ to $SO(n)$ upstairs, or from $S^{2n-1}\subset \C^n$ to 
$S^{n-1}\subset\R^n$ downstairs), and therefore 
to simplify  considerably the 
difficulty of the problem (see below for the reduction to real variables 
in the 3-level case).

In the present paper we take advantage of this procedure to find complete 
solutions to the optimal control problem for the three level quantum 
system \r{se}, \r{h-al} and the costs \r{energy-al}, \r{time-al}. More 
precisely:
\bd
\i[A.] the minimum time problem with
bounded controls 
(downstairs) is a problem in
dimension
five. In this case, since the dimension of the state space is
big, the problem of finding extremals and selecting optimal trajectories
can be extremely hard.
The fact that one can restrict to minimizers that are in
resonance permits to reduce the problem to a two-dimensional
problem, that can be solved with techniques similar to those used in 
\cite{libro}. This is the goal of Section \ref{s-mt};
\i[B.] the minimum energy problem  is 
naturally lifted  to a right invariant subriemannian problem on the group
$SU(3)$ (as explained in Remark \ref{r-lift}). This problem cannot be 
solved with the techniques used in
\cite{q2} for the isotropic case, because now the cost is built with a
``deformed Killing form''. Anyway since we can restrict to 
resonant minimizers the problem is reduced to a contact subriemannian
problem  on $SO(3)$, that does not have abnormal extremals (since it is
contact) and the corresponding Hamiltonian system is completely
integrable (since it leads to a right invariant Hamiltonian system on a 
Lie group of dimension 3, see for instance
\cite{jurd-book,jurd-MCT}). 
The complete solutions can be found in terms of  Elliptic 
functions. This is the aim of Section \ref{s-me}.  
\ed
\brem
Thanks to the reduction given in \cite{q3}, a 
minimum energy problem for
a  $n$-level system, with levels coupled by pairs, is a singular-Riemannian problem on $S^{n-1}$ or 
can be lifted to a right invariant subriemannian problem on $SO(n)$. 
Anyway there is no reason to believe that the Hamiltonian system 
associated to the PMP is Liouville integrable for $n\geq4$.
See \cite{sevilla} for some numerical solutions to this problem for $n=4$.
\erem

\subsection{Reduction to Real Variables}\llabel{s-rrv}

In the following, for the 3-level problem \r{se}, \r{h-al} we recall how 
to reduce the problem to real variables using the fact that we can restrict  
to resonant controls, i.e., controls of the form
\bqn
\F_{j}(t)=u_{j}(t)e^{i[(E_{j+1}-E_{j})t +\xi_{j}]},~~j=1,2,
\eqnl{res}
where $\xi_{j}\in[-\pi,\pi]$ are some phases and 
$u_{j}(.):[0,T]\to\R$ 
are the amplitudes of the lasers that should be determined.
Using the so called interaction picture, i.e., making the unitary 
transformation
\bqn \llabel{trans1}
\psi(t) \to U^{-1}(t)\psi(t), \mbox{ where } 
U(t):=diag(\e^{-iE_1},\e^{-iE_2},\e^{-iE_3}),
\eqn
and next making the transformation (to kill the 
phases) 
\bqn\llabel{trans2}
\psi(t)\to V^{-1}\psi(t), \mbox{ where } 
V:=  diag(1,e^{i(-\pi/2-\xi_1)}, e^{i(-\pi-\xi_1-\xi_2)}),
\eqn
the \sceq becomes
\bqn
\frac{d\psi(t)}{dt}=H(t)\psi(t).
\eqnl{se-no-i}
Notice that these transformations \underline{leave invariant the source} 
$|\psi_1|^2=1$ \underline{and the 
target}   $|\psi_3|^2=1$. Now, one can restrict equation \r{se-no-i} to 
reals 
(i.e., to the sphere 
$S^2$, i.e., $\psi(.):[0,T]\to\R^3$, $\psi_1^2+\psi_2^2+\psi_3^2=1$) by 
taking a real 
initial condition (the initial phase is arbitrary). Thus the drift 
is eliminated and the Hamiltonian belongs to 
$\mathfrak{so}(3)$
\bqn
H=\left(\ba{ccc} 0& -u_1&0\\ u_1 
&0&-\al u_2\\
0&\al u_2&0 \ea\right).
\eqnl{h-no-i}
For more details on these transformations, see \cite{q1,q2,q3}.
After reduction to real variables, the states one and three are 
represented 
respectively by the couples of points $(\pm1,0,0)^T$ and $(0,0,\pm1)^T$. 
And the costs in
which we are interested become: 
\bqn
&&\mbox{energy in fixed time:~~~} \int_0^T \left(u_1(t)^2+ 
u_2(t)^2\right)dt,\llabel{energy-u}\\ 
&&\mbox{time with bounded controls:~~~} T=\int_0^T 1~dt, 
\mbox{~~under the conditions $|u_1(t)|\leq 1$ and $|u_2(t)|\leq 
1$}.\llabel{time-u}  
\eqn
\brem
\llabel{r-positive}
Beside the proof that there always
exist minimizers in resonance, 
another useful result
given in \cite{q3} is that  (after reduction to real variables), 
\underline{there always exists a} 
\und{minimizer corresponding to positive}  \und{coordinates} 
(and it is also given the description of how to reconstruct
all other resonant minimizers from this last one). Hence we can 
assume 
$\psi\in  
S^+$ where $S^+$ is the subset of the 
sphere $S^2$ corresponding to 
positive coordinates
\bqn
S^{+}=\{\psi=(\psi_1,\psi_2,\psi_3)^T\in\R^3:~~
\psi_1^2+\psi_2^2+\psi_3^2=1,~~\psi_i\geq0
\}.
\eqnl{s+}
\erem
Let $Lip([0,T],S^+)$ be the set of Lipschitz functions from $[0,T]$ to $S^+$.
Our minimization problem, in which we precise the functional spaces, is:

\medskip\noindent
{\bf Problem downstairs (PD)} {\it Consider the control system 
\r{se-no-i}, 
\r{h-no-i}, 
where  $u_i(.)\in L^\infty([0,T],\R)$ ($i=1,2$), $\psi(.)\in 
Lip([0,T],S^+)$. 
Find the optimal trajectory and control steering the point 
$\psi=(1,0,0)^T$ (still called {source} or state one), 
to the point  $\psi=(0,0,1)^T$ (still called {target} or state 
three), that minimizes the cost \r{energy-u} or \r{time-u}.}
\brem
\llabel{r-opt}
We attack the problem by computing all trajectories in $S^+$, starting 
from  $(1,0,0)^T$, and  satisfying the PMP (the so called 
\underline{extremals}). Then we have to select, 
among all extremal trajectories reaching the target, those having 
smallest cost.  Since we will find that all extremals passing through a 
point of  $S^+\setminus \{(1,0,0)^T\}$ coincide before this point,
we get a stronger result than a solution to the problem 
{\bf (PD)}. Indeed, we obtain the complete \und{optimal synthesis}, i.e., 
a set of optimal trajectories starting from $(1,0,0)^T$ and reaching 
every point of $S^+$. For a more sophisticated definition of optimal synthesis,
see \cite{libro,piccoli-sussmann}. 
The optimal synthesis can be useful for applications, 
when one needs to reach a final state that is not an eigenstate.
\erem
For what concerns the problem upstairs, after the transformations \r{trans1}, \r{trans2} the equation for 
the operator of temporal evolution is
\bqn
\dot g(t)=Hg(t),
\eqnl{se-ev-no-i}
where now $g(t)\in SO(3)$ and $H$ is given by \r{h-no-i}. To pass from the 
problem upstairs to the 
problem downstairs, one should take the first column of $g(t)$, i.e., 
should
use the projection
\bqn
\ba{rcl}\Pi:SO(3)&\to& S^2\\
g&\mapsto& g\psi(0)=g(1,0,0)^T.
\ea
\eqnl{proj} 
Upstairs the source and the target are not points, but are 
respectively the sets $\Sur$ and $\Tur$ of matrices of SO(3) that projected with \r{proj} 
give rise to $(1,0,0)^T$ and $(0,0,1)^T$. More precisely we have
\bqn
\Sur&:=&\left(\mbox{
\begin{tabular}{c|c}
$1$&0 \\\hline
~&~\\
0&$~SO(2)~$ \\
~&~\\
\end{tabular}
}
\right)=SO(2)\subset SO(3),\llabel{s-up}\\
\Tur&:=&g_0~\Sur,~~~\mbox{where }
g_0:=\left(\ba{ccc}0&1&0\\0&0&1\\1&0&0\ea\right)\in SO(3),~~i.e.,~~ 
\Tur=\left(\mbox{
\begin{tabular}{c|c}
~&~ \\
0&$~SO(2)~$\\
~&~ \\
\hline
1&~0~
\end{tabular}
}
\right).
\llabel{t-up}
\eqn
Notice that $\Sur$ is a subgroup of $SO(3)$ while $\Tur$ is not (indeed it is a 
translation of a subgroup). The problem upstairs is finally:

\medskip\noi
{\bf Problem upstairs (PU)} {\it Consider the control system 
\r{h-no-i}, \r{se-ev-no-i}
where  $u_i(.)\in L^\infty([0,T],\R)$ ($i=1,2$), $g(.)\in 
Lip([0,T],SO(3))$. 
Find the optimal trajectories and controls connecting the source $\Sur$ to 
the target $\Tur$   
that minimize the cost \r{energy-u} or \r{time-u}.}
\brem
Notice that after elimination of the drift, both control systems 
\r{se-no-i}, 
\r{h-no-i}, and \r{h-no-i}, \r{se-ev-no-i} can be rewritten in the form (called 
\underline{distributional}) 
\bqn 
\dot y = u_1F_1(y) + u_2F_2(y),
\eqnl{distib}
where $y\in S^+$ downstairs,  or  $y\in SO(3)$ upstairs and 
\bqn
F_1(y)=
\left(\ba{ccc} 0&-1&0\\ 1
&0&0\\
0&0&0 
\ea\right)~y,~~~~
F_2(y)=\left(\ba{ccc} 
0&0&0\\ 
0&0&-\al\\
0&\al&0 
\ea\right)~y.
\eqnn
A problem with a dynamics \r{distib} and cost  \r{energy-u} belongs to the
frameworks of subriemannian geometry and singular-Riemannian geometry (see \cite{bellaiche,montgomery}).
\erem
Using standard arguments 
of controllability on compact Lie
groups and corresponding homogeneous spaces, one
gets the following:
\bp
After elimination of the drift term and reduction to real variables, 
the control system upstairs \r{h-no-i}, \r{se-ev-no-i} (resp. the 
control system downstairs \r{se-no-i}, \r{h-no-i})
is completely
controllable on $SO(3)$ (resp. on 
$S^+$).
\ep

\brem
Notice that both problems {\bf (PD)} and {\bf (PU)} 
can be seen as minimum time problems on a compact 
manifold with velocities in a convex compact set (and there is complete controllability).
Then, by Filippov theorem, they have solutions (see for instance 
\cite{cesari}).
\erem
\subsection{Structure of the Paper}
In Section \ref{s-PMP} we state PMP, our main tool, and we show 
that, for our minimization problems, there are no abnormal extremals. In 
Section \ref{s-mr} we state our main 
results.
In Section \ref{s-mt} we treat the minimum time problem downstairs. 
In Subsection \ref{s-2dman} we develop a theory of time optimal 
syntheses on 2-D manifolds for distributional systems with 
bounded 
controls (this subsection is written in such a way to be as 
self-consistent 
as possible)
and in the next subsection we build explicitly the time 
optimal synthesis.
In Section \ref{s-me}, attacking the problem upstairs, we treat the 
minimum energy problem. We give  explicit expressions of the optimal 
controls, for which the  corresponding trajectories form an optimal 
synthesis  on $S^+$.

\section{Pontryagin Maximum Principle}\llabel{s-PMP}
Our main tool is the Pontryagin Maximum Principle (PMP), stated below for 
the two cases in which the final time is free or  fixed.
For the proof see for instance \cite{agra-book,pontlibro}. In this 
version the source and the target are not just points, but more generally 
smooth submanifolds. This is useful while working upstairs (where the 
source  and the target are defined respectively by \r{s-up} and \r{t-up}).

\medskip
\noi{\bf Theorem (Pontryagin Maximum Principle)}
{\it Consider the control system $\dot x=f(x,u)$ with a
cost of the form $\int_0^Tf^0(x(t),u(t))~dt$, where the final time $T$ is 
\underline{free}, and initial and final
conditions given by $x(0)\in M_{in}$, $x(T)\in M_{fin}$,
where $x$
belongs to  a manifold $M$ and $u(.):[0,\infty[\to U\subset \R^m$. Assume moreover that
$M,~f,~f^0$ are smooth and that $M_{in}$ and $M_{fin}$ are smooth
submanifolds of $M$.

Define for every $(x,\lam,u)\in T^\ast M\times U$
\bqn
\HHH (x,\lam,u)&:=&<\lam, f(x,u)>+\lam_0f^0(x,u).\llabel{HAMHAMHAM}
\eqnn
If the couple $(x(.),u(.)):[0,T]\to M\times U$ (with $u(.)$ 
measurable and essentially bounded, and $x(.)$ Lipschitz) is 
optimal, then there
exists a \und{never vanishing} Lipschitz function 
$(\lam(.),\lam_0):t\in [0,T]\mapsto 
(\lam(t),\lam_0)\in
T^\ast_{x(t)}M\times \R$
(where $\lam_0\leq0$ is  a constant) such that for a.e. $t\in[0,T]$ we have:
\begin{description}
\item[i)]
$\dot x(t)=$ {\large $\frac{\partial \HHH
}{\partial \lam}$}$(x(t),\lam(t),u(t))$,
\item[ii)] $\dot \lam(t)=-${\large$\frac{\partial \HHH
}{\partial x}$}$(x(t),\lam(t),u(t))$,
\item[iii)] $\HHH (x(t),\lam(t),u(t))=\HHH_M(x(t),\lam(t))$, where 
$\HHH_M(x,\lam):=\max_{v\in U}\HHH(x,\lam,v).$
\item[iiii)] $\HHH_M(x(t),\lam(t))=0.$
\item[v)] $<\lam(0),T_{x(0)}M_{in}>=<\lam(T),T_{x(T)}M_{fin}>=0$ 
(transversality
conditions).
\end{description}
If the final time $T$ is \underline{fixed by hypothesis}, then condition 
{\bf 
iiii)} 
must be replaced by 
\bd
\item[iiiibis)] $\HHH_M(x(t),\lam(t))=k(T)\geq0$, for some constant 
$k(T)$.
\ed
}
\bdeff
The map $\lam(.):[0,T]\to T_{x(t)}^\ast M$ is called \underline{covector}.  
The real-valued map on $T^*M\times U$, defined in \r{HAMHAMHAM} is called
\underline{PMP-Hamiltonian}. A trajectory $x(.)$ (resp. a couple
$(x(.),\lam(.))$) satisfying conditions
{\bf i)},
{\bf ii)},
{\bf iii)} and
{\bf iiii)} (or {\bf iiiibis)}) is called an \underline{extremal} (resp. 
an 
\underline{extremal
pair}).
If $(x(.),\lam(.))$ satisfies {\bf i)},
{\bf ii)},
{\bf iii)} and
{\bf iiii)}  (or {\bf iiiibis)}) 
with $\lam_0=0$ (resp. $\lam_0<0$), then it is called an
\underline{abnormal extremal} (resp. a \underline{normal extremal}). 
\edeff
\brem
Notice that the definition of abnormal extremal does not depend on the
cost but only on the dynamics (in fact if $\lam_0=0$, the cost disappears 
in \r{HAMHAMHAM}). For general properties of abnormal extremals, we refer to
\cite{bonnard-book,montgomery}.
\erem

\bp
\llabel{p-no-ab}
For the control systems downstairs \r{se-no-i}, \r{h-no-i}, there are no 
abnormal extremals. The 
same holds for the control system upstairs \r{h-no-i}, \r{se-ev-no-i}. 
\ep
\proof 
For the problems  \r{se-no-i}, \r{h-no-i} and \r{h-no-i}, 
\r{se-ev-no-i}, the set of admissible 
velocities at a point $x$ writes
$
\{u_1F_1(x)+u_2F_2(x)~|~(u_1,u_2)\in U\}, 
$
where the set $U$ is a subset of $\R^2$ 
containing 0 in its interior. This implies that, if there were an
abnormal extremal, 
its covector $\lambda(.)$ should annihilate $F_1$ and $F_2$ all along the 
trajectory, 
but also, using PMP, their bracket $[F_1,F_2]$. But for both problems {\bf (PD)} and {\bf (PU)}, 
$(F_1,F_2,[F_1,F_2])$ 
forms a generating family of the tangent space which implies that 
$\lambda(.)\equiv 0$, all 
along the trajectory. This is forbidden by the PMP.
\quadp

\section{Main Results}\llabel{s-mr}
In this section we present explicitly \underline{the optimal controls steering 
state one to state three} for our two costs. The \underline{optimal 
syntheses}
are presented in Sections \ref{s-mt} and \ref{s-me} and pictured in Figures \ref{f-synt} and 
\ref{f-nonnepossopiu}.
In both cases, all the extremals issued from $(1,0,0)^T$ given by the PMP 
are 
optimal in $S^+$ and they leave $S^+$ in finite time. For the minimum 
energy problem,  there is no cut-locus, i.e., two optimal trajectories 
starting from the source at time zero,  
cannot intersect for positive time.
On the other side, for the minimum time problem, two optimal trajectories 
can intersect for $t>0$, but in this case they coincide before 
the intersection.
\brem
Notice that due to the symmetries of the problem, for both costs, if 
$(u_1(.),u_2(.)):[0,T]\to\R\times\R$ are the optimal 
controls steering the state one to the state three and corresponding to a 
nonisotropy factor $\al$, then $(\bar u_1(.),\bar u_2(.)):[0,\al T[\to 
\R\times\R$, defined by $(\bar u_1(t),\bar 
u_2(t)):=(u_2(T-t/\al),u_1(T-t/\al)),$
are the optimal controls steering the state one to the state three 
and corresponding to a nonisotropy factor $1/\al$.  
\erem
\subsection{Minimum Time}\llabel{TimeMainResult}
For the minimum time problem, optimal controls are a.e. piecewise 
constant, and the corresponding
trajectory is a concatenation of arcs of circle.
There are three different cases depending  on the value of $\al$: 
\bd
\i[\underline{Case $0<\al<1$}.]
In this case the minimizer corresponds to the control:
\bi
\item $u_1(t)=u_2(t)=1$ for $t\in[0,\frac{\arccos(-\al^2)}{\sqrt{1+\al^2}}[$. 
The corresponding trajectory steers the point $(1,0,0)^T$ to the point
$(0,\sqrt{1-\al^2},\al)^T$.
\item $u_1(t)=0,~u_2(t)=1$ for $t\in[\frac{\arccos(-\al^2)}{\sqrt{1+\al^2}},
\frac{\arccos(-\al^2)}{\sqrt{1+\al^2}}+\frac{\arccos{\al}}{\al}]$. 
The corresponding trajectory steers the point $(0,\sqrt{1-\al^2},\al)^T$ 
to 
the point $(0,0,1)^T$.
\ei 
\i[\underline{Case $\al=1$}.]
In this case the minimizer corresponds to $u_1(t)=u_2(t)=1$ for $t\in[0,
\frac{\pi}{\sqrt{2}}]$ and steers the state one to the state three.
\i[\underline{Case $\al>1$}.]
In this case the minimizer corresponds to:
\bi
\item $u_1(t)=1,~u_2(t)=0$ for $t\in[0,\arccos(\frac{1}{\al})[$. 
The corresponding trajectory steers the point $(1,0,0)^T$ to the point 
$(\frac{1}{\al},\sqrt{1-\frac{1}{\al^2}},0)^T$.
\item $u_1(t)=u_2(t)=1$ for $t\in[\arccos(\frac{1}{\al}),
\arccos(\frac{1}{\al})+
\frac{1}{\sqrt{1+\al^2}}\arccos(-\frac{1}{\al^2})]$. 
The corresponding trajectory steers the point 
$(\frac{1}{\al},\sqrt{1-\frac{1}{\al^2}},0)^T$ to the point $(0,0,1)^T$.
\ei 
\ed
\subsection{Minimum Energy}

For the minimum energy problem, there exists a unique strictly positive 
number $m_3(0)$ such that the trajectory, corresponding to the controls 
given below, steers the state
$(1,0,0)^T$ to the state $(0,0,1)^T$ and  is 
parameterized by  arclength. The parameter $m_3(0)$, appearing in these 
formulas, should be computed numerically, for instance by the
dichotomy method presented in Section \ref{s-numerical-shit}.

\bd
\i[\underline{Case $0<\al\leq1$}.]
$u_1(t)=\mbox{\emph{cn}}
         \left ( \al ~m_3(0) ~t;k \right ),~~
         u_2(t)=  \mbox{ \emph{sn}} \left (
         \al ~m_3(0) ~t;k \right ),$ 
          where
         we have defined the modulus of the elliptic functions $cn$ and
         $sn$ to be equal to
        $k=\frac{\sqrt{1-\al^2}}{\al~ m_3(0)}.$

\i[\underline{Case $\al>1$}.] $u_1(t)=\mbox{\emph{cd}}
         \left ( \frac{\sqrt{\al^2-1}}{k}~t;k \right ),
         u_2(t)= \sqrt{1-k^2}~\mbox{\emph{sd}} \left (
         \frac{\sqrt{\al^2-1}}{k}
          ~t;k \right ),$ 
         where
         we have defined the modulus of the elliptic functions $cd$ and
         $sd$ to be equal to
         $k=\sqrt{\frac{\al^2-1}{\al^2 m_3(0)^2 + \al^2 -1}}.$
\ed
For the parameter $m_3(0)$ we have the following estimate:
$  \sqrt{\frac{1-\al^2}{\al^2}}  < m_3(0)\leq \sqrt{\frac{4}{3 
\al^2}-1}$,  if $0<\al\leq 1$, and   $0<m_3(0)\leq \frac{1}{\sqrt{3}}$, if 
$\al> 1$.
The time needed to reach the target is given by formula \r{Tm1} if 
$\al\leq 1$ and by formula  \r{Tp1}  if $\al> 1$.
In the particular case $\al=1$, we recover the results of 
\cite{q2}. 
\section{Minimum Time}
\llabel{s-mt}
\subsection{Minimum Time for Distributional Systems on 2-D 
Manifolds}
\llabel{s-2dman}
In this section, we develop a theory of time
optimal
syntheses on 2-D manifolds for distributional systems with bounded
controls. For this purpose we use ideas
similar to those used by Sussmann, Bressan,
Piccoli and the first author in
\cite{automaton,quattro,due,sus1,sus2}
for the minimum time stabilization to the origin for
the ``control affine version'' of the same problem
($\dot x=F(x)+uG(x)$, $u\in[-1,1]$, where $x$ belongs to a
2-D-manifold) and recently
rewritten in
\cite{libro}.
This section is written to be as self-consistent as 
possible.
\subsubsection{Basic Definitions and PMP}
We focus on the following:

\medskip\noi
{\bf Problem (P)} {\it
Consider the control system  
\bqn \dot x = u_1F_1(x) + u_2F_2(x),~~~x\in M,~~~|u_i|\leq 1,~~i=1,2,
\eqnl{mt}
where
\bd
\i[(H0)] $M$ is a smooth 2-D manifold. The vector fields 
$F_1$ and $F_2$ are 
$\con^\infty$ and the control system \r{mt} is complete on $M$. 
\ed
We are interested in the problem of \und{reaching every point 
of $M$ in minimum time} from a source $M_{in}$ that is a smooth 
submanifold of  
$M$.}\\\\
The theory developed next is then applied to {\bf (PD)}:
\bqn 
\left\{\ba{l}
\dot \psi = u_1F_1(\psi) + u_2F_2(\psi),~~~\psi\in S^+,~~~|u_i|\leq 
1,~~i=1,2,\\
F_1(\psi)=
\left(\ba{ccc} 0&-1&0\\ 1
&0&0\\
0&0&0 
\ea\right)
\left(\ba{c}\psi_1\\\psi_2\\\psi_3\ea\right)=
\left(\ba{c}-\psi_2\\\psi_1\\0\ea\right)=\displaystyle-\psi_2
\frac{\partial}{\partial 
\psi_1}+\psi_1\frac{\partial}{\partial \psi_2},\\
F_2(\psi)=\left(\ba{ccc} 
0&0&0\\ 
0&0&-\al\\
0&\al&0 
\ea\right)
\left(\ba{c}\psi_1\\\psi_2\\\psi_3\ea\right)=
\al\left(\ba{c}0\\-\psi_3\\\psi_2\ea\right)=\displaystyle\al\left(-
\psi_3\frac{\partial}{\partial 
\psi_2}+\psi_2\frac{\partial}{\partial \psi_3}\right).
\ea\right.
\eqnl{mts}
\bdeff
A control for the system \r{mt} is a measurable function 
$u(.)=(u_1(.),u_2(.)):[a_1,a_2]\to 
[-1,1]^2$. The 
corresponding trajectory is a Lipschitz continuous map
$x(.):[a_1,a_2]\to M$ such that $\dot 
x(t)=u_1(t)F_1(x(t))+u_2(t)F_2(x(t))$ for
almost every $t\in [a_1,a_2]$. Since the system is autonomous we 
can always
assume that $[a_1,a_2]=[0,T]$. 
\edeff
For us, a solution to the problem {\bf (P)} is an \und{optimal synthesis} 
that is a  
collection $\{(x_{\bar x}(.), u_{\bar x}(.))$ defined on $[0,T_{\bar 
x}], \bar x\in M\}$ of trajectory--control pairs
such that $x_{\bar x}(0)\in M_{in}$, $x_{\bar x}(T_{\bar x})=\bar x$, and 
\und{$x_{\bar x}(.)$ 
is time 
optimal}.

In the following we use the notation $u=(u_1,u_2)$ and (in a local chart) 
$x=(x_1,x_2)$, $F_1=((F_1)_1,(F_1)_2)$, 
$F_2=((F_2)_1,(F_2)_2)$.  
Let us introduce a definition to describe different types of controls. 
\bdeff
Let $u(.)=(u_1(.),u_2(.)):[a_1,a_2]\subset[0,T]\to[-1,1]^2$ be a 
control for 
the 
control system \r{mt}.
\bi 
\i $u(.)$ is 
said to be a \und{bang control} if for almost
every  $t\in[a_1,a_2]$, $u(t)=\bar u\in 
\{(-1,-1),(-1,1),(1,-1),(1,1)\}.$ Similarly $u(.)$ is 
said to be a \und{$u_i$-bang control} if for almost
every  $t\in[a_1,a_2]$, $u_i(t)=\underline u\in\{\pm1\}.$

\i A \und{switching}  time of
$u(.)$ is a time $t\in[a_1,a_2]$ such that for every $\eps>0$, $u(.)$  is 
not bang on $(t-\eps,t+\eps)\cap[a_1,a_2]$. Similarly 
a \und{$u_i$-switching}  time of
$u(.)$ is a time $t\in[a_1,a_2]$ such that for every $\eps>0$, $u(.)$  
is not $u_i$-bang on $(t-\eps,t+\eps)\cap[a_1,a_2]$. 

\i  If $u_A:[a_1,a_2]\to[-1,1]^2$ and 
$u_B:[a_2,a_3]\to[-1,1]^2$ are 
controls, their \underline{concatenation} $u_B\ast u_A$ is the
control
$$
(u_B\ast u_A)(t):=\left\{\ba{l}
u_A(t)\mbox{ for } t\in[a_1,a_2],\\
u_B(t)\mbox{ for } t\in]a_2,a_3].\ea\right.
$$ 
The control $u(.)$ is called  \und{bang-bang} if 
it is a finite  concatenation of bang arcs. Similarly one defines 
\und{$u_i$-bang-bang} controls.

\i A trajectory of \r{mt} is a \und{bang trajectory},
(resp. \und{bang-bang trajectory}),  if it corresponds
to a bang control, (resp. bang-bang control). Similarly one defines 
$u_i$-bang and $u_i$-bang-bang trajectories.

\ei
\edeff
A key role is played by the following three functions defined on 
$M$
\bqn 
\Delta_A(x)&:=&Det(F_1(x)),F_2(x))= (F_1)_1(F_2)_2-(F_2)_1(F_1)_2,
\llabel{deltaA}\\
\Delta_{B_1}(x)&:=&Det(F_1(x)),[F_1,F_2](x))= 
(F_1)_1([F_1,F_2])_2-([F_1,F_2])_1(F_1)_2,\llabel{deltaB1}\\
\Delta_{B_2}(x)&:=&Det(F_2(x)),[F_1,F_2](x))=  
(F_2)_1([F_1,F_2])_2-([F_1,F_2])_1(F_2)_2\llabel{deltaB2}.
\eqn
Notice that these definitions depend  on the choice of the coordinate 
system, but not  the sets $\da,$ $\dbu,$ $\dbd$ of their zeros, that 
are respectively the set of points where $F_1$ and
$F_2$ are parallel, the set of points where $F_1$ is parallel to
$[F_1,F_2]$ and
the set of 
points where $F_2$ is parallel to 
$[F_1,F_2]$. Using PMP it turns out (see Section \ref{s-stps}) that 
these loci are fundamental in the construction of the optimal 
synthesis.
In fact,
assuming that they are embedded one-dimensional submanifolds of $M$,
we have the following:
\bi
\i in each connected region of $M\setminus(\da\cup \dbu\cup\dbd)$, every 
extremal
trajectory is bang-bang
with at most two switchings (one of the control $u_1$ and one of 
the control $u_2$). Moreover the possible switchings are determined. More 
precisely for every $x\in M\setminus(\da\cup \dbu\cup\dbd)$ define 
\bqn
f_i(x):=-\frac{\Delta_{B_i}(x)}{\Delta_A(x)}.
\eqnl{fsw}
If $f_i>0$ (resp. $f_i<0$), we have that $u_i$ can only 
switch from $-1$ to $+1$ (resp. from $+1$ to $-1$);

\i the support of  $u_i$-singular trajectories (that are trajectories for 
which the $u_i$-switching function identically vanishes, and for which 
$u_i$ can assume values different from $\pm1$, see Definition 
\ref{d-sw-f} below) is always
contained in the set $\dbi$.
\ei
For the problem {\bf (P)}, the PMP  says the following:

\medskip
\noi{\bf Corollary (Pontryagin Maximum Principle for the problem (P))}
{\it Consider the  control system \r{mt} subject to {\bf (H0)}.
For every $(x,\lam,u)\in T^\ast M\times [-1,1]^2$, define
\bqn
\HHH(x,\lam,u)&:=&u_1<\lam,F_1(x)>+u_2<\lam,F_2(x)>+\lam_0,
\eqnn
If the couple $(x(.),u(.)):[0,T]\to M\times [-1,1]\times [-1,1]$
is time optimal then there exist a \underline{never vanishing} 
Lipschitz continuous \und{covector} 
$\lam(.):t\in[0,T]\mapsto \lam(t)\in 
T^\ast_{x(t)}M$ and a constant $\lam_0\leq 0$ such that for a.e. $t\in [0,T]$:
\begin{description}
\item[i)]
$\dot x(t)=$ {\large $\frac{\partial \HHH
}{\partial \lam}$}$(x(t),\lam(t),u(t))$,
\item[ii)] $\dot \lam(t)=-${\large$\frac{\partial \HHH
}{\partial x}$}$(x(t),\lam(t),u(t))=-<\lam(t),(u_1(t)\nabla 
F_1+u_2(t)\nabla F_2)(x(t))>$,
\item[iii)] $\HHH (x(t),\lam(t),u(t))=\HHH_M(x(t),\lam(t))$, ~~where~~ 
$\HHH_M(x,\lam):=\max\{\HHH(x,\lam,u): u\in [-1,1]^2\},$
\item[iiii)] $\HHH_M(x(t),\lam(t))=0$,
\item[v)] $<\lam(0),T_{x(0)}M_{in}>=0$ 
(transversality condition).
\ed
}
\brem\llabel{r-nedjma}
In this version of PMP, $\lam(.)$ is always different from zero otherwise 
the conditions {\bf iii)}, {\bf iiii)} would imply $\lam_0=0$ (cf. PMP in 
Section \ref{s-PMP}).  An extremal is said to be \underline{nontrivial} if 
it  does not correspond to controls a.e. vanishing. Notice that a trivial 
extremal is an abnormal extremal.
In the following we  often refer to \underline{nontrivial abnormal 
extremal} (\und{NTAE}, for short). 
\erem
\subsubsection{Switching Functions, Singular Trajectories and Predicting 
Switchings}
\llabel{s-stps}
In this section we  are interested
in determining when the controls switch from $+1$  to $-1$ or 
viceversa and when they may assume values in $]-1,+1[$. Moreover we  
would like to predict which kind of switchings can happen, using 
properties of the vector fields $F_1$ and $F_2$.  A key role is played by 
the following:
\bdeff {\bf (Switching Functions)}
Let $(x(.),\lam(.))$ be an extremal pair. The corresponding
switching functions are defined as $\phi_i(t):=<\lam(t),F_i(x(t))>$, 
$i=1,2$.
\llabel{d-sw-f}
\edeff
\brem
Notice that $\phi_i(.)$ are at least Lipschitz continuous. 
Moreover using the
switching functions, conditions  {\bf iii)} and 
{\bf iiii)} imply:
\bqn
\HHH(x(t),\lam(t),u(t))=u_1(t)\phi_1(t)+u_2(t)\phi_2(t)+\lam_0=0\mbox{ 
a.e.}.
\eqnl{hphi}
\erem
The following lemma characterizes abnormal extremals.
\bl 
\llabel{l-ab}
Let $(x(.),\lam(.))$ (defined on $[a_1,a_2]$) be an extremal 
pair. We have:\\
{\bf 1.} $(x(.),\lam(.))$ is an abnormal extremal if and only if  
$\phi_1(.)\equiv\phi_2(.)\equiv 0$ on $[a_1,a_2]$;\\
{\bf 2.} if $(x(.),\lam(.))$ is  an abnormal extremal, then 
$Supp(x(.))\subset\da$;\\ 
{\bf 3.} if  $(x(.),\lam(.))$ is not an abnormal extremal, then  
$\phi_1(.)$ 
and $\phi_2(.)$ never vanish at 
the same time;\\ 
{\bf 4.} let $\bar t\in[a_1,a_2]$ be a time such that $\phi_1(\bar t)=0$, 
$x(\bar t)\in\da$, $F_1(x(\bar t))\neq0$. Then $(x(.),\lam(.))$ is  an 
abnormal extremal. 
The same holds if $\phi_2(\bar t)=0$,
$x(\bar t)\in\da$, $F_2(x(\bar t))\neq0$.
\el
\proof Let us prove {\bf 1.} The sufficiency is obvious. Let 
us prove the necessity. Equation \r{hphi}, with $\lam_0=0$ and {\bf iii)} 
of PMP, imply that both controls vanish (trivial abnormal extremals) or that
$\phi_1(.)\equiv\phi_2(.)\equiv0$ on $[a_1,a_2]$.
Let us prove {\bf 2.} From $\phi_1(.)\equiv\phi_2(.)\equiv0$ we have 
that $\lam(.)$ is orthogonal to $F_1(x(.))$ and $F_2(x(.))$. Since 
$\lam(.)$ is never vanishing, it follows that $F_1(x(.))$ is parallel to 
$F_2(x(.))$.
To prove {\bf 3,} assume by contradiction that  $(x(.),\lam(.))$ is 
not an abnormal extremal
and that there exists a time $\bar t$ for which 
$\phi_1(\bar t)=\phi_2(\bar t)=0$. From \r{hphi} there exists a sequence 
$t_m\nearrow \bar t$ such that $u_1(t_m)$ and $u_2(t_m)$ are defined and
$u_1(t_m)\phi_1(t_m)+u_2(t_m)\phi_2(t_m)+\lam_0=0$. Since 
$\phi_i(t_m)\to0$, $i=1,2$, it follows $\lam_0=0$. Contradiction.
To prove {\bf 4.}, observe that $\phi_1(\bar t)=0$ implies that $\lam(\bar 
t)$ annihilates $F_1(x(\bar t))$. From $x(\bar t)\in \da$ we have 
that 
$F_2(x(\bar t))$ is parallel to $F_1(x(\bar t))$, hence $\lam(\bar t)$ 
annihilates also $F_2(x(\bar t))$. It follows $\phi_2(\bar t)=0$. From 
\r{hphi} we have $\lam_0=0$. The same proof is valid exchanging
the two indexes 1 and 2. 
\quadp

\noi
The following lemma restricts the set where the support of abnormal 
extremals can live. 
\bl
\llabel{l-highlynon}
Let $(x(.),\lam(.))$  be an abnormal extremal. Then 
$Supp(x(.))\subset 
\da\cap\dbu\cap\dbd$.  
\el
\proof
Let $x(.)$ defined on $[a_1,a_2]$ be an abnormal extremal. Then 
condition {\bf 
2.} of Lemma \ref{l-ab} implies that $F_1$ and $F_2$ are parallel 
along $Supp(x(.))$. It 
follows that the distribution restricted to $Supp(x(.))$ is integrable in 
the  Frobenius sense. Hence also $[F_1,F_2]$ is parallel to $F_1$ and 
$F_2$ along 
$Supp(x(.))$. By definition of $\Delta_{B_i}$ it follows that $\Delta_A$, 
$\Delta_{B_1}$, $\Delta_{B_2}$ are zero along $x(.)$. \quadp

\noi
We recall that for the problem \r{mts}, there are no abnormal extremals 
(cf. Proposition \ref{p-no-ab}). Anyway, as a consequence of Lemma 
\ref{l-highlynon},  
the  presence of NTAE is highly nongeneric  also for the general system 
\r{mt} subject to  {\bf (H0)}, since generically, points of 
$\da\cap\dbu\cap\dbd$ are isolated.

\medskip\noi
The switching functions determine when the controls switch from $+1$ to 
$-1$ 
and viceversa. In fact, from 
the maximization condition {\bf iii)}, one immediately gets:
\bl\llabel{l-sw}
Let $(x(.),\lam(.))$ defined on $[0,T]$ be an extremal 
pair and $\phi_i(.)$ the
corresponding switching functions. If $\phi_i(t)\neq 0$ for some
$t\in]0,T[$,
then there exists $\eps>0$
such that $x(.)$ corresponds to a constant control $u_i=sgn(\phi_i)$  on
$]t-\eps,t+\eps[$.
Moreover if $\phi_i(.)$ has
a zero at $t$, and if $\dot \phi_i(t)$ exists and  is strictly larger 
than
zero (resp. strictly smaller than zero) then
there exists
$\eps>0$
such that $x(.)$ corresponds to constant control $u_i=-1$  on
$]t-\eps,t[$ and to  constant control  $u_i=+1$  on
$]t,t+\eps[$ (resp. to  constant control  $u_i=+1$  on
$]t-\eps,t[$ and to  constant control  $u_i=-1$  on
$]t,t+\eps[$).
\el
Notice that on every
interval where
$\phi_i(.)$ has no zero (resp. finitely many zeroes) the
corresponding control is $u_i$-bang (resp. $u_i$-bang-bang).
We are then interested in differentiating $\phi_i$. 
One immediately gets:
\bl\llabel{l-dif-phi}
Let $(x(.),\lam(.))$, defined on $[0,T]$ be an extremal pair
and $\phi_i(.)$
the corresponding switching functions. Then it holds 
a.e.
$\dot\phi_1(t)=u_2(t)<\lam(t),[F_2,F_1](x(t))>,$
$\dot\phi_2(t)=u_1(t)<\lam(t),[F_1,F_2](x(t))>.$
\el
From Lemma \ref{l-sw} it follows that $u_i$ can assume values different 
from $\pm1$ on some interval $[a_1,a_2]$ only if the corresponding 
switching 
function vanishes identically there.
\brem
Lemma \ref{l-ab} asserts that, if there are no NTAE (as for the system 
\r{mts}),  then $u_1$ and $u_2$ never switch at
the same time. In this case,  from Lemma  \ref{l-dif-phi} it follows that 
in a \neigh of 
a $u_1$-switching,
$\phi_1(.)$ is a $\con^1$ function. A similar statement holds for 
$\phi_2(.)$.
\erem
\bdeff
A nontrivial extremal trajectory $x(.)$  defined on $[a_1,a_2]$
is said to be \underline{$u_i$-singular}
if the corresponding switching function
$\phi_i(.)$ vanishes identically on $[a_1,a_2]$.
\edeff
\brem
From Lemma \ref{l-ab} it follows that $(x(.),\lam(.))$ 
is a NTAE if and only if it is a $u_1$-$u_2$-singular 
trajectory. 
\erem
Next we show that an extremal
trajectory, between two $u_i$-switchings, intersects the set 
$\da\cup\dbu\cup\dbd$. Moreover we prove that 
$u_i$-singular 
trajectories must run on the set $\dbi$. Let us define two 
crucial 
functions.
\bdeff
\llabel{c2:def-f}
On the set of points $x\in M\setminus\da$, the vector fields  
$F_1(x),F_2(x)$ form a
basis of $T_{x}M$ and we define
the scalar functions $f_1(x),f_2(x)$ to be
the coefficients of the linear combination:
$[F_1,F_2](x)=f_2(x)F_1(x)-f_1(x)F_2(x)$.
\edeff
The following Lemma gives a relation 
between $f_1$, $f_2$, 
and the functions $\Delta_A$, $\Delta_{B_1}$, $\Delta_{B_2}$.
\bl
Let $x\in M\setminus\da$ then $\displaystyle
f_i(x)=-\frac{\Delta_{B_i}(x)}{\Delta_A(x)},$ $i=1,2.$
\el
\proof
We have $\Delta_{B_1}(x)=Det\big(F_1(x),[F_1,F_2](x)\big)=Det(F_1(x),f_2(x)F_1(x)-
f_1(x)F_2(x))=$\\$=-f_1(x) Det\big( F_1(x), 
F_2(x)\big)=-f_1(x)\Delta_A(x),$
and similarly for $\Delta_{B_2}$.
\quadp\\\\
The functions $f_1$ and $f_2$  are crucial in studying which kind of 
switchings can happen near $u_i$-ordinary points defined next:
\bdeff
\llabel{d-ord}
A point $x\in M$ is called a \underline{$u_i$-ordinary
point} if $x\notin\da\cup\dbi$. 
\edeff
On the set of $u_i$-ordinary points the structure of optimal trajectories
is particularly simple:
\bp
\llabel{cap2:s1}
Let $\Omega\subset M$ be an open connected set made of 
$u_i$-ordinary points. Then all extremal trajectories 
$x(.):[a_1,a_2]\to \Omega$,  
are $u_i$-bang-bang with at most one  $u_i$-switching.
Moreover if $f_i>0$ (resp. $f_i<0$)
in $\Omega$
then $x(.)$ corresponds to control $u_i$ that is:\\
{\bf $\bullet$} a.e. constantly equal to  $+1,$ or\\
{\bf $\bullet$}  a.e. constantly equal to $-1$ or\\  
{\bf $\bullet$}  has a $-1\to +1$ switching (resp. has a $+1\to -1$ 
switching).
\ep
\proof
Let $x(.):~]a_1,a_2[\to\Omega$ be an extremal trajectory and 
$\phi_1(.)$ be the corresponding $u_1$-switching function.
If $\phi_1(.)$ has no zero, then $x(.)$ is a $u_1$-bang and the 
conclusion follows. 
Let $t_1$ be a zero of $\phi_1(.)$. 
The time  $t_1$ cannot be a zero of $\phi_2(.)$ otherwise $x(t_1)$ could 
not be a $u_i$-ordinary point (by Lemma \ref{l-ab} we would have 
$\Delta_A(x(t_1))=0$). From Lemma \ref{l-dif-phi} it 
follows that $\phi_1(.)$ is $\con^1$ in a \neigh of $t_1$.
Moreover  $t_1$ 
cannot be a zero of $\dot \phi_1(.)$ otherwise $x(t_1)$ could
not be a $u_1$-ordinary point (we would have $\Delta_{B_1}(x(t_1))=0$).
Since in a
\neigh of $t_1$,
$u_2$ is a.e. constantly equal to $+1$ or $-1$, we can assume $u_2$ 
constant in this neighborhood, and we have
\bqn
\dot\phi_1(t_1)&=&u_2(t_1)<\lam(t_1),[F_2,F_1](x(t_1))>=
u_2(t_1)<\lam(t_1),(-f_2 F_1+f_1 
F_2)(x(t_1))>\nn\\
&=&u_2(t_1)f_1(x(t_1))~<\lam(t_1),F_2(x(t_1))>=
u_2(t_1)f_1(x(t_1))~\phi_2(t_1).
\eqn
Now from Lemma \ref{l-sw} we have that $sgn(u_2(t_1))=sgn(\phi_2(t_1))$, 
and since $\phi_2(t_1)\neq0$ it follows  $sgn(\dot 
\phi_1(t_1))=sgn(f_1(x(t_1)))$. 
Using again Lemma \ref{l-sw}, it follows that if $f_1>0$ (resp. 
$f_1<0$) then we can have only a $-1\to+1$ switching (resp.
a $+1\to-1$ switching). 
A similar proof can be done for a zero of $\phi_2$. \quadp

\medskip\noi
We are now interested in properties of $u_i$-singular 
trajectories.
\bl
\llabel{l-deltaB}
Let $(x(.),\lam(.))$, defined on $[0,T]$, be a $u_i$-singular trajectory 
on 
$[a_1,a_2]\subset[0,T]$,  
 then $Supp(x(.)|_{[a_1,a_2]})\subset \dbi$.
\el
\proof
If $(x(.),\lam(.))$ is a NTAE, the conclusion follows from Lemma 
\ref{l-highlynon}. Assume now that  $(x(.),\lam(.))$ is not a NTAE.
To simplify the notation assume $i=1$, the case $i=2$ being similar. From 
$\phi_1(.)=<\lam(.),F_1(x(.))> 
\equiv0$ on $[a_1,a_2]$ 
we have that $\dot \phi_1(.)=u_2(.)<\lam(.),[F_2,F_1](x(.))>\equiv0$ a.e. 
in  $[a_1,a_2]$. 
Since $(x(.),\lam(.))$ is not a NTAE, it follows that $u_2$ is a.e. equal 
to $+1$ or $-1$ in $[a_1,a_2]$. Therefore $\lam(.)$ is orthogonal 
both to $F_1(x(.))$ and 
$[F_1,F_2](x(.))$. Being $\lam(.)$ nontrivial then 
$F_1(x(.))$ 
is parallel to 
$[F_1,F_2](x(.))$. The conclusion follows. \quadp 
\bl
\llabel{l-varphi}
Let $(x(.),\lam(.))$ be a $u_1$-singular trajectory (resp. 
a $u_2$-singular 
trajectory) on $[a_1,a_2]$.
Assume  $<\nabla\Delta_{B_1}(x(t)),F_1(x(t))>\neq0$ (resp. 
$<\nabla\Delta_{B_2}(x(t)),F_2(x(t))>\neq0$) and define respectively
\bqn
\varphi_1(t):=-u_2(t){<\nabla \Delta_{B_1}(x(t)),F_2(x(t))>\over 
<\nabla\Delta_{B_1}(x(t)),F_1(x(t))>},~~
\varphi_2(t):=-u_1(t){<\nabla \Delta_{B_2}(x(t)),F_1(x(t))>\over 
<\nabla\Delta_{B_2}(x(t)),F_2(x(t))>}.
\eqnn
Then $\varphi_1(.)$ (resp. 
$\varphi_2(.)$) satisfies
$|\varphi_1(t)|\leq1$ (resp. $|\varphi_2(t)|\leq1$) 
a.e. in
$[a_1,a_2]$. 
Moreover $x(.)$ 
corresponds to the control $(\varphi_1(.),u_2(.))$ (resp. 
$(u_1(.),\varphi_2(.))$ in 
$[a_1,a_2]$.
\el
\proof
Assume $i=1$, the case $i=2$ being similar. Let $u_1(.)$, $u_2(.)$  be the
controls corresponding to $x(.)$, that is
$\dot x(t)=u_1(t)F_1(x(t))+u_2(t)F_2(x(t))$, for almost every $t$. 
From $\Delta_{B_1}(x(t))=0$, we have for a.e. $t$:
$0=\frac{d}{dt}\Delta_{B_1}(x(t))=\nabla\Delta_{B_1}(x(t))\cdot(u_1(t)F_1(x(t))
+u_2(t)F_2(x(t))).$
This means that $u_1(t)=\varphi_1(t)$. The condition $|\varphi_1(t)|\leq1$ 
is simply a  consequence of the fact that,  by assumption, $x(.)$ is 
an admissible trajectory.
\quadp
\brem
Notice that the three functions $\Delta_A$, $\Delta_{B_1}$,  
$\Delta_{B_2}$ are not invariant 
by change of coordinates and/or orientation of $M$. Anyway, the sets of 
their zeros, the functions $f_i$ (see Definition 
\ref{c2:def-f}) and the functions $\varphi_i$ (see Lemma \ref{l-varphi}) do 
not depend on the 
coordinate system.  
\erem
\subsubsection{More on Singular Trajectories}
From Lemma \ref{l-deltaB} it follows that the condition 
$\dbi\neq\emptyset$ is  
necessary for the presence of a $u_i$-singular trajectory. Anyway  
in general it is not sufficient. It may 
very well happen that  $\dbi\neq\emptyset$ and no singular trajectory is 
running on it. Another common situation is that a singular trajectory 
is running just a subset of $\dbi$. 

The description of singular trajectories is  very intricate in general 
(see for instance \cite{libro} for the control affine case). Here we 
describe what happens in the simplest case (that is enough for what 
follows), i.e., when we restrict our attention to a 
subset 
$S_1$ of $\dbu$ satisfying the following 
conditions:
\bd
\i[(C1)] $S_1$ is a smooth one-dimensional connected 
embedded submanifold 
of $M$, 
\i[(C2)] $S_1$ does not intersect  $\da$ and $\dbd$, 
\i[(C3)] $S_1$ is not tangent to the vectors $F_1+F_2$ and
$-F_1+F_2$,
\i[(C4)] $f_1$ changes sign on $S_1$.
\ed
See Figure \ref{f-mista-1} A. Similarly, we restrict our attention to a 
subset  $S_2$ of $\dbd$ satisfying the same
conditions {\bf (C1)}--{\bf (C4)}, but with the indexes $1$ and $2$ 
exchanged.
\ppotR{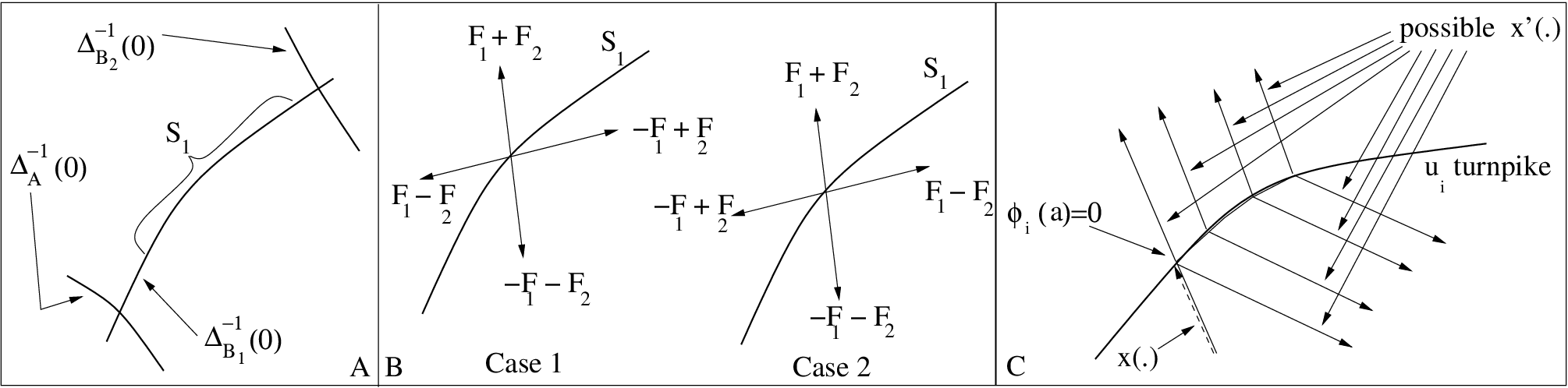}{Hypotheses on $\dbu$ (Figure A). Possible positions 
of the vector 
fields $F_1$ and $F_2$ on $\dbu$  (Figure B). Shape of the synthesis in a 
\neigh of a point where a trajectory is becoming $u_i$-singular (Figure 
C)}{16}

Let us treat the case of $S_1$, being the case of $S_2$ similar.
First observe that, if a $u_1$-singular trajectory defined in 
$[a_1,a_2]$ is running on $S_1$, then $u_2$ cannot switch on $[a_1,a_2]$, 
otherwise, by Lemma \ref{l-ab}, $x(.)$ would be a NTAE (and this is not the case thanks to 
{\bf (C2)}).  Assume, for 
instance, $u_2=1$.
Due to {\bf (C3)}, $F_1+F_2$ and $-F_1+F_2$ point either on the opposite, 
or on the same sides of $S_1$ (see Figure \ref{f-mista-1} B, cases 1 and 2). 
Since the admissible velocities belong to the segment joining $F_1+F_2$ 
and $-F_1+F_2$ it follows that, only in case 1 of Figure \ref{f-mista-1} 
B, 
it is possible to run on $S_1$. 
Moreover, using a similar argument to that of Lemma 12 p. 47 of 
\cite{libro}, one can prove that if 
$f_1<0$ on the side where $F_1+F_2$ points, 
then a trajectory 
running on $S_1$ cannot be optimal. 
These facts are collected in the following: 
\bl
Let $S_1$ (resp. $S_2$) be a subset of $M$ satisfying conditions 
{\bf (C1)},
{\bf (C2)},
{\bf (C3)}, {\bf (C4)}  above. Then on $S_1$ (resp. on $S_2$) cannot run 
an optimal $u_1$-singular trajectory (resp. an optimal $u_2$-singular 
trajectory) if (at least) one of the two conditions is 
satisfied.
\bd
\i[OO1] $F_1+F_2$ and $-F_1+F_2$ (resp. $F_1+F_2$ and $F_1-F_2$)
point on the same side of $S_1$ (resp. $S_2$);
\i[OO2] $F_1+F_2$ and $-F_1+F_2$ point on opposite sides of $S_1$ (resp. 
$F_1+F_2$ and $F_1-F_2$ point on opposite sides of $S_2$), and  in a 
\neigh of  $S_1$ (resp. $S_2$) we have the following. On 
the side where points $F_1+F_2$, we have $f_1<0$ (resp. 
$f_2<0$).
\ed
\el
If on $S_1$ (resp. on $S_2$)  {\bf OO1} and {\bf OO2} are not satisfied 
(that means that $S_1$ is a candidate support for an optimal trajectory)
then, following \cite{libro}, we call $S_1$ a \underline{$u_1$-turnpike} 
(resp. a \underline{$u_2$-turnpike}).

An extremal trajectory $x(.)$, corresponding to control $u_1$ constantly 
equal to $+1$ or $-1$ and  
reaching a $u_1$-turnpike at a point $\bar x= x(\bar t)$, can enter it only if 
the corresponding $u_1$-switching function is vanishing  at $\bar t$, 
otherwise, 
after $\bar t$, it will correspond to the same control. 
Moreover, using an argument similar to that of Lemma 11 p. 46 of 
\cite{libro} one can prove that if an extremal trajectory enters a 
turnpike, then it can exit it with control $+1$ or $-1$ at every 
successive 
time. These facts are stated in the following Lemma, and 
illustrated in Figure \ref{f-mista-1} C.
\bl
\llabel{l-entra-esci}
Let $(x(.),\lam(.))$ defined on $[0, a]$, $a>0$, be  a bang  extremal pair 
that
verifies $x(a)=\bar x$, $\bar x\in S_i$ where $S_i$ is a $u_i$-turnpike,
and assume that $\phi_i(a)=<\lam(a),F_i(x(a))>=0$.
Moreover let $b,c$ be two real numbers, sufficiently close to $a$,  such 
that 
$a\leq b< c$, and let
$x'(.):[0, c]\to M$ be a trajectory, corresponding to controls 
$u'_1(.)$ and $u'_2(.)$ such that:
\bi
\i $x'(.)|_{[0,a]}=x(.)$,
\i  $Supp(x'(.)|_{[a,b]})\subset S_i$, and $u'_j(.)$, $j\neq i$, is bang 
on 
$[0,c]$,
\i $x'(.)|_{]b,c]}$ is bang.
\ei
Then $x'(.)$ is extremal. Moreover, if $\phi'_i(.)$ is the $u_i$-switching
function
corresponding to $x'(.)$, then $\phi'_i|_{[a,b]}\equiv 0$.
\el
\brem
For the problem {\bf (P)}, under generic conditions on the vector fields 
$F_1$ and $F_2$, one can classify  synthesis 
singularities, stable syntheses, singularities of the minimum time wave 
fronts, with the same techniques used in \cite{libro}.
\erem
\subsection{Time Optimal Synthesis for the 3-Level Quantum System}
In this section, we apply the theory developed in Section \ref{s-2dman} to 
the system \r{mts}.
Here our source is the point $(1,0,0)^T$ and the task is to reach the 
target $(0,0,1)^T$ in minimum time. 
As explained in Remark \ref{r-opt}, 
the first step is to compute all 
extremal trajectories in $S^+$.  Then we have to select,
among all extremal trajectories reaching the target, the one having
smallest cost.  
Since it turns out that all these 
trajectories 
are  optimal (because all extremals passing through a
point of  $S^+\setminus \{(1,0,0)^T\}$ coincide before this point)
it follows global
optimality, and we get the complete time optimal 
synthesis in $S^+$.

When one is dealing with a minimum time problem for a linear system $\dot 
x=Ax+Bu$, $x\in\R^2$, $A\in\R^{2\times2}$, $B\in\R^2$, $u\in[-1,1]$, the 
equation for $\lam(.)$, given by PMP, 
is decoupled from the equation for $x(.)$ and does not contain the control, 
hence for every value of  $\lam(0)$ one can compute $\lam(t)$ and the 
corresponding switching  function $<\lam(t),B>$ from which one determines the value of 
the control. For our problem \r{mts}, as consequence of  
bilinearity, the  equation for $\lam(.)$ 
does not depend on $x(.)$, but still contains the control.  For this 
reason  we have to use the theory developed 
before, that uses the functions
$\Delta_A$, $\Delta_{B_i}$, $f_i$.

One of the most important points is to prove that there is no chattering 
(i.e., an infinite concatenation of bang  and/or 
singular arcs in finite time). In synthesis theory this step is called the 
\und{finite dimensional 
reduction} because it permits to restrict the set of candidate optimal 
trajectories to a family of trajectories parameterized by a finite number 
of parameters (the switching times). 
The absence of chattering is proved in Proposition 
\ref{p-pa-chat}. Thanks to that, we will be able to prove that every 
optimal control
is piecewise constant and this permits to solve explicitly the equation for $\lam$ 
(see Section \ref{s-cs}).

\subsubsection{Preliminary Computations and Facts}
\llabel{s-prel}
From \r{mts} we have in the standard coordinates of 
$\R^3$
\bqn
[F_1,F_2](\psi)=\al \left( -\psi_3\parti{\psi_1}+ 
\psi_1\parti{\psi_3}\right),~~~
\Da  = \al \psi_2,~~~
\Dbu  =  \al \psi_1,~~~
\Dbd =  -\al^2 \psi_3.
\eqnn
Hence
\bqn
\da  =  \{\psi\in S^+~|~\psi_2=0\},~~~
\dbu  =  \{\psi\in S^+~|~\psi_1=0\},~~~
\dbd  =  \{\psi\in S^+~|~\psi_3=0\}.
\eqnn
They form the boundary of $S^{+}$. Using Lemma 
\ref{l-deltaB}, the two sets $\dbu,~\dbd$  
permit to locate singular trajectories (if they exist).
Now we want to compute 
the values of the singular controls, i.e., the 
functions $\varphi_1$ and $\varphi_2$ defined 
in Lemma \ref{l-varphi}. They are identically 
equal to zero where they are defined. Indeed $\Dbu=\al \psi_1$, hence   
$F_2$ is orthogonal to $\nabla\Dbu$ on $\dbu$. Notice that
$\varphi_1$ is not defined 
when $F_1$ is also orthogonal to the gradient of $\Dbu$, 
(hence when $F_1$ and $F_2$ are parallel, that is along $\da$),  
anyway at these points $\varphi_1$ can be defined by continuity.
A similar computation works for $\varphi_2$.
These facts are collected in the following:
\bp \llabel{p-zero}
If an extremal trajectory $x(.)$ is  $u_1$-singular (resp. $u_2$-singular) 
in $[a_1,a_2]\subset[0,T]$, then, in 
$[a_1,a_2]$, it corresponds to 
control $u_1$ (resp. $u_2$) equal to zero and its 
support belongs to $\dbu$ (resp. $\dbd$), i.e., the circle of 
equation $\psi_1=0$ (resp. $\psi_3=0$).
\ep
Now let us study which switchings are admitted. In $S^+$ we have
\bqn
f_1(\psi) & = & -\frac{\psi_1}{\psi_2}<0,\mbox{ where 
}\psi_2\neq0,~\psi_1\neq0,\nn\\
f_2(\psi) & = & \al \frac{\psi_3}{\psi_2}>0,\mbox{ where
}\psi_2\neq0,~\psi_3\neq0.\nn
\eqn
\brem
Using the sign of these two functions one immediately checks that in $S^+$ 
the set  $\dbu\setminus\{(0,1,0)^T,(0,0,1)^T\}$ is a $u_1$-turnpike and 
the set   
$\dbd\setminus\{(1,0,0)^T,(0,1,0)^T\}$
is a $u_2$-turnpike.
\erem
Hence using Proposition \ref{cap2:s1} we get:
\bp
\llabel{p-int}
In $S^+\setminus(\da\cup\dbu)$ (resp. $S^+\setminus(\da\cup\dbd)$) every 
extremal trajectory has at most one $u_1$-switching  (resp. one 
$u_2$-switching) according to the following rules:
$u_1 : +1\longrightarrow -1, ~~(\mbox{resp. } u_2 : -1\longrightarrow 
+1).$
\ep
\begin{figure}
\begin{center}
\input{f-leq.pstex_t}
\llabel{f-leq}
\caption{The vector fields $F_1$, $F_2$, $[F_1,F_2]$, the sets $\da$, $\dbu$, $\dbd$, 
and the functions $f_1$ and $f_2$ on $S^+$.}
\end{center}
\end{figure}
Since there are no abnormal extremals, using {\bf 4.} of Lemma \ref{l-ab}, 
one 
gets the next proposition, that is  useful in the following to prove 
that there is no chattering.
\bp
\llabel{p-pa-deltaA}
Let $(x(.),\lam(.))$ defined on $[0, T]$ be  an extremal 
pair. Let 
$\bar t\in[0,T]$ be such that $x(\bar t)\in\da\setminus\{(0,0,1)^T\}$ 
(resp. $x(\bar t)\in\da\setminus\{(1,0,0)^T\}$). Then 
$\phi_1(x(\bar t))\neq0$ (resp. $\phi_2(x(\bar t))\neq0$).
\ep
The next proposition shows that \underline{every extremal is a 
finite concatenation of bang and singular
arcs}.
\bdeff
Let $(x(.),\lam(.))$ defined on $[0,T]$ be an extremal pair. A time  
$\bar t\in [0,T]$ is said to be a \underline{chattering time} 
if there exists a nontrivial  sequence $t_n$, tending to $\bar t$, with 
$u_1$ or $u_2$ being not bang or singular  on any \neigh of $t_n$.
\edeff
\bp
\llabel{p-pa-chat}
Let $(x(.),\lam(.))$ defined on $[0,T]$ be an extremal pair. 
Then no $\bar t\in[0,T]$ is  a {chattering time}.
\ep
\proof 
Let us prove the proposition for  $u_2$, being the proof for $u_1$ 
equivalent.
Assume by contradiction that there exists a time $\bar t\in[0,T]$ of 
$u_2$-chattering and that the chattering is on the 
left of $\bar t$, being the opposite case equivalent.
By Proposition \ref{p-int}, $x(\bar t)$ is not in 
$S^+\setminus(\da\cup\dbd)$. Moreover since  $\phi_2(\bar t)=0$, by 
Proposition \ref{p-pa-deltaA} we have that $x(\bar t)$ does not belong to 
$\da\setminus\{(1,0,0)^T\}$. Hence $x(\bar t)\in \dbd$.

We claim that there are two times $\tau_1$ and $\tau_2$ such that
$0<\tau_1<\tau_2<\bar t$, $x(\tau_1)\in \dbd$ and $x(\tau_2)\notin \dbd$.
Indeed, if it is not the case, then on some interval $]\bar t-\eps,\bar t[$ 
($\eps>0$), the support of $x(.)$ is included in $\dbd$ or in $S^+\setminus\dbd$
and in both cases there is no chattering. 
As a consequence of the form of the dynamics (see Figure \ref{f-synt}), 
in $S^+$, an extremal can leave
and come back to $\dbd$ only having a $+1\to-1$ switching of $u_2$ in 
$S^+\setminus\dbd$ which is forbidden by Proposition \ref{p-int}.
\quadp
\brem\llabel{r--10}
Thanks to Propositions \ref{p-zero} and \ref{p-pa-chat}, \underline{every extremal trajectory 
corresponds to piecewise constant} \underline{controls}, i.e., 
\underline{it is a finite concatenation of arcs of circle}.
\erem

\brem\llabel{r--11}
Once the control $u_1$ (resp. $u_2$) takes value $-1$ (resp. 1) on a nontrivial 
interval of time, it will not switch anymore in $S^+$. Indeed, outside 
$\dbu$ 
(resp. $\dbd$) it cannot switch (see Proposition \ref{p-int}), and the 
trajectory cannot reach $\dbu$ (resp. $\dbd$) with control $u_1=-1$ 
(resp. $u_2=1$).
\erem

\subsubsection{Construction of the Synthesis} \llabel{s-cs}
In the following we construct the complete time optimal synthesis. 
First we write a (linear) system of equations for the switching functions
and its solutions along a bang or singular arc, that can be easily verified:  

\bp \llabel{p-ode}
Let $(x(.),\lam(.))$ defined on $[0,T]$ be an extremal pair. 
Recall that $\phi_i(.)=<\lam(.),F_i(x(.))>$, $i=1,2$ and define 
$\phi_3(.):=<\lam(.),[F_1,F_2](x(.))>$.  Then, along a bang or a singular 
arc, the triplet $(\phi_1,\phi_2,\phi_3)$ is solution of the time invariant 
ODE
\bqn
\left\{
\ba{l}
\dot\phi_1=-u_2\phi_3\\
\dot\phi_2=u_1\phi_3\\
\dot\phi_3=\al^2u_2\phi_1-u_1\phi_2,
\ea
\right.
\eqnl{phi123}
i.e., we have
\bqn
\left(
\ba{c}
\phi_1(t)\\
\phi_2(t)\\
\phi_3(t)
\ea
\right)
= R(t)
\left(
\ba{c}
\phi_1(0)\\
\phi_2(0)\\
\phi_3(0)
\ea
\right),
\eqnl{mat123}
where
\bqn
R(t)=
\left(
\ba{ccc}
\frac{{u_1}^2 + 
     \al^2\,{u_2}^2\,\cos (t\,
        {\sqrt{{u_1}^2 + \al^2\,{u_2}^2}})}{{u_1}^2 + 
     \al^2\,{u_2}^2} & - \frac{u_1\,u_2\,
       \left( -1 + \cos (t\,{\sqrt{{u_1}^2 + 
               \al^2\,{u_2}^2}}) \right) }{{u_1}^2 + 
       \al^2\,{u_2}^2}   & - \frac{u_2\,
       \sin (t\,{\sqrt{{u_1}^2 + \al^2\,{u_2}^2}})}{{\sqrt{{
             u_1}^2 + \al^2\,{u_2}^2}}}   \cr -
     \frac{\al^2\,u_1\,u_2\,
       \left( -1 + \cos (t\,{\sqrt{{u_1}^2 + 
               \al^2\,{u_2}^2}}) \right) }{{u_1}^2 + 
       \al^2\,{u_2}^2}   & \frac{\al^2\,{u_2}^2 + 
     {u_1}^2\,\cos (t\,{\sqrt{{u_1}^2 + 
            \al^2\,{u_2}^2}})}{{u_1}^2 + \al^2\,{u_2}^2}
   & \frac{u_1\,\sin (t\,
       {\sqrt{{u_1}^2 + \al^2\,{u_2}^2}})}{{\sqrt{{u_1}^2 + \al^2\,
       {u_2}^2}}} \cr \frac{\al^2\,u_2\,
     \sin (t\,{\sqrt{{u_1}^2 + \al^2\,{u_2}^2}})}{{\sqrt{{
           u_1}^2 + \al^2\,{u_2}^2}}} & - \frac{
       u_1\,\sin (t\,{\sqrt{{u_1}^2 + 
             \al^2\,{u_2}^2}})}{{\sqrt{{u_1}^2 + 
          \al^2\,{u_2}^2}}}   & \cos (t\,
    {\sqrt{{u_1}^2 + \al^2\,{u_2}^2}}) 
\ea
\right).
\eqnl{mat1234}

\ep

\noindent Since there is no chattering, we have:

\bp
For every extremal trajectory $x(.)$ such that 
$x(0)=(1,0,0)^T$, there exists $\eps>0$ such that it
corresponds to controls $(u_1,u_2)=(1,0)$, or to control 
$(u_1,u_2)=(1,1)$, in $[0,\eps]$.
\ep
\proof
Since $(1,0,0)^T\notin\dbu$, $u_1$ cannot be zero for small times. 
Moreover $u_1=-1$ or $u_2=-1$
do not permit to  $x(.)$ to enter $S^+$.   \quadp

\brem
Notice that, since $F_1$, $F_2$ and $[F_1,F_2]$ is a generating family of the tangent space,
the knowledge of $\lam$ is equivalent to the knowledge of $(\phi_1,\phi_2,\phi_3)$.
Since the initial covector $\lam(0)$ is free, one have some freedom on the choice
of $(\phi_1(0),\phi_2(0),\phi_3(0))$. More precisely, starting from the point $(1,0,0)^T$, we have $\phi_2(0)=0$ (because $F_2=0$),  
$\phi_1(0)>0$ (otherwise $u_1$ would not be 1 for small time) and $\phi_3(0)$ is free. Moreover, since 
the covector is defined up to a positive constant, we can normalize $\phi_1(0)=1$.

Notice also that a $u_1$-singular (resp. $u_2$-singular) extremal defined on $[a,b]$ has
$\phi_1(.)=\phi_3(.)=0$ (resp. $\phi_2(.)\equiv\phi_3(.)\equiv0$) on $[a,b]$.
\erem

In the following propositions, we study in details the two extremals starting 
from $(1,0,0)^T$ and corresponding to controls $(1,0)$ and $(1,1)$ and the
possible switchings along them. Notice that if 
$\al<1$ then the trajectory corresponding to controls $u_1=u_2=1$ 
intersects the boundary of $S^+$ on $\dbu$ while if $\al>1$   it intersects  
the boundary of $S^+$ on $\da$. 

\bp\llabel{p-10local}
The trajectory starting from $(1,0,0)^T$ and corresponding 
to control $u_1=+1$, $u_2=0$  is 
extremal up to time $\pi/2$ (i.e., until it leaves $S^+$). Moreover for 
every $a\in]0,\pi/2[$ there exists $\eps>0$  such that the trajectory 
starting from $(1,0,0)^T$ and corresponding 
to control $(1,0)$ in $[0,a]$ and to control $(1,1)$ in $]a,a+\eps]$ is 
extremal.
\ep

\proof The first claim is immediately checked by plugging $u_1=1$, $u_2=0$
inside the equations \r{mat123} and \r{mat1234} for time $t\in[0,\frac{\pi}{2}]$, 
and choosing initial covector in such a way that $\phi_1(0)=1$ and 
$\phi_2(0)=\phi_3(0)=0$. Since along this extremal the $\phi_i(.)$ are constant it
follows that the trajectory starting from $(1,0,0)^T$ and corresponding 
to control $u_1=+1$, $u_2=0$  is extremal until it leaves $S^+$.

The second claim can be seen as a consequence of Lemma \ref{l-entra-esci}.
Anyway, it can be checked directly by setting $\phi_1(0)=1$, 
$\phi_2(0)=\phi_3(0)=0$ and plugging $u_1=1$, $u_2=0$ inside the equation  \r{mat123} and \r{mat1234} 
for time $t\in[0,a]$ and  $u_1=1$, $u_2=1$ for time $t\in]a,a+\eps]$.
Finally one have to check that $\phi_1(.)$ and $\phi_2(.)$ remain positive  for an $\eps$ 
small enough.
\quadp

\noindent Similarly one gets the following:

\bp\llabel{p-11local}
The trajectory starting from $(1,0,0)^T$ and corresponding 
to control $u_1=+1$, $u_2=+1$ is 
extremal up to time $T_\al$ where
\bqn
T_\al=
\left\{
\ba{l}
\frac{\arccos(-\al^2)}{\sqrt{1+\al^2}} \mbox{ if } \al\leq 1\\
\frac{1}{\sqrt{1+\al^2}}\arccos(-\frac{1}{\al^2})  \mbox{ if } \al \geq 1.
\ea
\right.
\eqnl{t-al}
(Notice that if $\al\leq 1$, $T_\al$ is the time in which the curve reaches $\dbu$
and leaves $S^+$.)
Moreover for every $0<a< T_\al$ 
there exists $\eps>0$ sufficiently small such that the trajectory 
starting from $(1,0,0)^T$ and corresponding 
to controls $(u_1,u_2)=(1,1)$ in $[0,a]$ and to controls $(u_1,u_2)=(-1,1)$ in $]a,a+\eps]$ is 
extremal. The same happens if $a=T_\al$ and $\al\neq 1$.
\ep
Next we call $\g^{++}$ defined on $[0,T_\al]$, the extremal trajectory 
starting from $(1,0,0)^T$ and corresponding to controls $(u_1,u_2)=(1,1)$.
In the two following propositions we prolong most of the extremals built in Propositions
\ref{p-10local} and \ref{p-11local}.
\bp\llabel{s-t-c}
Let $(x(.),\lambda(.))$ be an extremal pair starting from $(1,0,0)^T$, 
corresponding to controls $(u_1,u_2)=(1,0)$ in the interval $[0,a]$ ($0<a<\frac{\pi}{2}$) and 
switching to controls $(1,1)$ at time $a$. Then:\\
{\bf A.} If $\al\leq 1$, there is no other switching before leaving 
$S^+$.\\
{\bf B.} If $\al>1$ and $a\geq \arccos(\frac{1}{\al})$, there is no other 
switching 
before leaving $S^+$.\\
{\bf C.} If $\al>1$ and $a< \arccos(\frac{1}{\al})$, then the control 
$u_1$ switches to $-1$ at time
$a+T_\al$ and there is no other switching before leaving $S^+$.
\ep
\proof
First remember that after time $a$, the control $u_2$ does not switch (see Remark \ref{r--11}).
Since $x(.)$ is $u_2$-singular in $[0,a]$ and corresponds to control 
$u_1=1$, we have
$\phi_1(a)=1$ and $\phi_2(a)=\phi_3(a)=0$. Using equation \r{mat123} and \r{mat1234} one easily
checks when $\phi_1$ changes of sign. This allows to check if there is a $u_1$-switching before
leaving $S^+$.
Moreover remember that once $(u_1,u_2)=(-1,1)$ then there is no additional
switching (see Remark \ref{r--11}). This allows to conclude for the last claim.
\quadp

\brem
Notice that in case {\bf C.} the set of points where the extremal trajectories 
are switching from controls $(1,1)$ to $(-1,1)$ describes a curve that is a great 
arc of circle between the points $\g^{++}(T_\al)$ and the target $(0,0,1)^T$. Indeed this is a
consequence of the fact that the time between $\dbd$ (which is a great arc of circle) and 
the $u_1$-switching is constant. 
\erem
Similarly one gets:
\bp\llabel{s-t-nc}
Let $(x(.),\lambda(.))$ be an extremal pair, starting from $(1,0,0)^T$, corresponding to 
controls $(1,1)$ in $[0,a]$ with $a\leq T_\al$ and switching to controls $(-1,1)$ at time $a$.
Then it is extremal until it leaves $S^+$ by reaching $\da$. 
\ep

The following proposition concludes the construction of all extremal trajectories. The proof is
similar to the proofs above and it is left to the reader.
\bp
In the case $\al<1$, for every $a\in[T_\al, T_\al+T'_\al]$ where 
$T'_\al=\frac{\arccos(\al)}{\al}$ 
the trajectory  starting from $(1,0,0)^T$ and corresponding to controls:\\
$\bullet$ $(u_1,u_2)=(1,1)$ for $t\in[0,T_\al],$\\
$\bullet$ $(u_1,u_2)=(1,0)$ for $t\in]T_\al, a[,$\\
$\bullet$ $(u_1,u_2)=(1,-1)$ for $t\in[a, b]$, where $b$ is the last time 
in which the trajectory 
stays in $S^+$, \\ 
is extremal.
\ep
The previous propositions describe all the possible extremals starting from $(1,0,0)^T$. They are
pictured in Figure \ref{f-synt}. One immediately checks the following:\\\\
{\bf Claim:} Each point of $S^+$ is reached by an
extremal trajectory. Moreover,  all the extremal trajectories reaching a 
fixed point, coincide before
that point. This allows to conclude that \underline{all these extremals are 
in fact optimal} and we have
built the complete \underline{time optimal synthesis} for the system \r{mts} starting from $(1,0,0)^T$.

\medskip\noi
Notice that the only points reached by more than one extremal 
are on 
$Supp(\g^{++})$ and on the support of  singular trajectories.   These 
facts are collected in the following:

\bt 
The set of trajectories starting from $(1,0,0)^T$ and corresponding to the controls given
below is a time optimal synthesis for the system \r{mts}.

In the sequel we denote by $T$ the last time in which the control is defined,
that is the last time in which the corresponding trajectory stays in $S^+$.

\noindent{\bf Case $\al<1$}
$$
\begin{array}{l}
(u_1,u_2)=\left\{
\ba{l}
(1,0) \mbox{ for } t\in[0,a] \mbox{ with }a\in[0,\frac{\pi}{2}],\\
(1,1) \mbox{ for } t\in]a,T].
\ea
\right.\\
(u_1,u_2)=\left\{
\ba{l}
(1,1) \mbox{ for } t\in[0,a] \mbox{ with }a\in]0,\frac{\arccos(-\al^2)}{\sqrt{1+\al^2}}],\\
(-1,1) \mbox{ for } t\in]a,T].
\ea
\right.\\
(u_1,u_2)=\left\{
\ba{l}
(1,1) \mbox{ for } t\in[0,\frac{\arccos(-\al^2)}{\sqrt{1+\al^2}}],\\
(0,1) \mbox{ for } t\in]\frac{\arccos(-\al^2)}{\sqrt{1+\al^2}},a] \mbox{ with }
a\in]\frac{\arccos(-\al^2)}{\sqrt{1+\al^2}},\frac{\arccos(-\al^2)}{\sqrt{1+\al^2}}+
\frac{\arccos(\al)}{\al}],\\
(-1,1) \mbox{ for } t\in]a,T].
\ea
\right.\\
\mbox{The curve reaching the target } (0,0,1)^T \mbox{ corresponds to following controls:}\\
(u_1,u_2)=\left\{
\ba{l}
(1,1) \mbox{ for } t\in[0,\frac{\arccos(-\al^2)}{\sqrt{1+\al^2}}],\\
(0,1) \mbox{ for } t\in]\frac{\arccos(-\al^2)}{\sqrt{1+\al^2}},\frac{\arccos(-\al^2)}{\sqrt{1+\al^2}}+
\frac{\arccos(\al)}{\al}].
\ea
\right.\\
\ea
$$
{\bf Case $\al=1$}
$$
\ba{l}
(u_1,u_2)=\left\{
\ba{l}
(1,0) \mbox{ for } t\in[0,a] \mbox{ with }a\in[0,\frac{\pi}{2}],\\
(1,1) \mbox{ for } t\in]a,T].
\ea
\right.\\
(u_1,u_2)=\left\{
\ba{l}
(1,1) \mbox{ for } t\in[0,a] \mbox{ with }a\in[0,\frac{\pi}{\sqrt{2}}],\\
(-1,1) \mbox{ for } t\in]a,T].
\ea
\right.\\
\mbox{The curve reaching the target } (0,0,1)^T \mbox{ corresponds to following controls:}\\
(u_1,u_2)=\left\{
\ba{l}
(1,1) \mbox{ for } t\in[0,\frac{\pi}{\sqrt{2}}].
\ea
\right.\\
\ea
$$
\noindent{\bf Case $\al>1$}
$$
\begin{array}{l}
(u_1,u_2)=\left\{
\ba{l}
(1,0) \mbox{ for } t\in[0,a] \mbox{ with }a\in[\arccos(\frac{1}{\al}),\frac{\pi}{2}],\\
(1,1) \mbox{ for } t\in]a,T].
\ea
\right.\\
(u_1,u_2)=\left\{
\ba{l}
(1,0) \mbox{ for } t\in[0,a] \mbox{ with }a\in]0,\arccos(\frac{1}{\al})[,\\
(1,1) \mbox{ for } t\in]a,a+T_\al],\\
(-1,1) \mbox{ for } t\in]a+T_\al,T].
\ea
\right.\\
(u_1,u_2)=\left\{
\ba{l}
(1,1) \mbox{ for } t\in[0,a] \mbox{ with }a\in]0,T_\al],\\
(-1,1) \mbox{ for } t\in]a,T].
\ea
\right.\\
\mbox{The curve reaching the target } (0,0,1)^T \mbox{ corresponds to following controls:}\\
(u_1,u_2)=\left\{
\ba{l}
(1,0) \mbox{ for } t\in[0,\arccos(\frac{1}{\al})],\\
(1,1) \mbox{ for } t\in]\arccos(\frac{1}{\al}),\frac{1}{\sqrt{1+\al^2}}\arccos(-\frac{1}{\al^2})].
\ea
\right.\\
\ea
$$
\et

From the controls given above, one can easily get the explicit expression of the corresponding optimal
trajectories that are concatenations of at most three arcs of circle that are bang or 
singular. More details about optimal trajectories reaching the target 
$(0,0,1)^T$ are given
in Section \ref{TimeMainResult}.
\begin{figure}
\begin{center}
\input{f-synt.pstex_t}
\caption{Image of the synthesis for minimum time with bounded controls in the cases $\al<1$, 
$\al=1$ and $\al>1$}
\llabel{f-synt}
\end{center}
\end{figure}
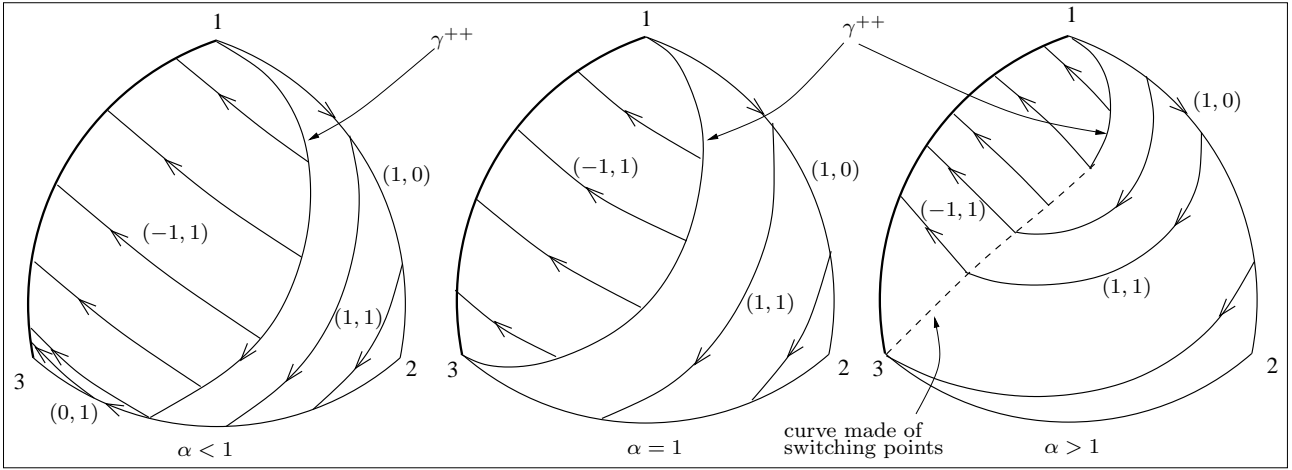



\section{Minimum Energy}
\llabel{s-me}

In this section, for the minimum energy problem,
we provide explicit expressions (in terms
of elliptic functions) for optimal controls linking the first and
the third level, in terms of a parameter that will be determined numerically.
In the following \underline{{we restrict our attention to 
extremal trajectories  parametrized by arclength}}.
This is possible thanks to 
the fact that,  in this special case,  the PMP-Hamiltonian system 
is Liouville integrable (since it is a right invariant Hamiltonian system 
on SO(3), 
see Section \ref{s-integr}). 
Some efforts are required to prove that all extremals are in fact optimal (see Section 
\ref{SEC-OptimalSynthesisSubRiem}).  Similarly to Section \ref{s-mt}, as 
byproduct we get the complete optimal synthesis (cf. Remark \ref{r-opt}).
In the sequel, for convenience of computations, we make a change of variables on the controls 
in such a way that the nonisotropy factor $\al$ does not appear in the
dynamics,  but in the cost.

\subsection{Change of Variables}

  A problem whose dynamics verifies an equation of the form of
  \r{h-no-i}, (\ref{se-ev-no-i}) is said to be \underline{right invariant}
  on the matrix group $SO(3)$. This group is endowed with a very special
  structure of real semi-simple compact Lie group. 

  In the following it is more natural to present the problem
  upstairs (i.e., at the level of the group), and in a slightly different 
way. Actually, we make 
  the change  of variables: $v_1(.)=u_1(.), ~~~v_2(.)=\al  u_2(.).$ 
In this way, at the level of the group $SO(3)$, 
the minimization problem becomes:
\bd
\i[Problem ${\cal P}$:] 
\bqn  
&&  \frac{dg}{dt}(t)
= \left ( \begin{array}{ccc}
             0& -v_1 & 0 \\
            v_1 & 0 & -v_2 \\
             0 & v_2 & 0
       \end{array}
      \right ) g(t) = dR_{g(t)}  \left ( \begin{array}{ccc}
             0& -v_1 & 0 \\
            v_1 & 0 & -v_2 \\
             0 & v_2 & 0
       \end{array}
      \right ),\label{EQ-dynsystem}\\
&&\mbox{minimize }{J}(v_1(.),v_2(.))=\int_0^T (v_1^2(t) + \frac{1}{\al^2} 
v_2^2(t))~dt.\label{EQ-cost}\\ 
&&g(0)\in{\cal S},~~~g(T)\in{\cal T}.
\eqn
\ed
Here $v_i:[0, T] \rightarrow \R$, for $i=1,2$ are measurable essentially  
bounded functions,
 and $dRg$ denotes the differential of  $R_g$,   
where $R_g$ is the right 
translation by $g$, i.e., 
$R_g g'=g'g$, where $g,g'\in SO(3)$.
Notice that 
$dRg$ is still a right 
translation by $g$. 
The cost function depending only on the controls 
(i.e., it is invariant under the group action)  is called right invariant.
The source ${\cal S}$ (i.e., the lift of the state one), and the target 
${\cal T}$  (i.e., the lift of the state three), are defined in 
\r{s-up}, \r{t-up}.
Recall that the  first column
$(\psi_1(t),\psi_2(t),\psi_3(t))^T$  of $g(t)$ represents the position (on the sphere $S^2$) of the system downstairs (cf. Formula 
\r{proj}).
Recall moreover that, in the minimization \r{EQ-cost}, 
to guarantee the existence of minimizers, the final time
$T$ should be fixed. In the following, we fix $T$ in such a way 
that the trajectories  are parameterized by
arclength.

\subsection{Integration of the Hamiltonian System Given by the PMP}
\llabel{s-integr}

   For the problem ${\cal P}$ the Lie group structure allows a
 nice formulation of the Pontryagin Maximum Principle. To take advantage of this formulation, 
 we need some
 extra notions, presented in the next paragraph. What follows can be found
  in a more general setting with proofs and further discussions
  in \cite{agra-book} or \cite{jurd-book}.

\subsubsection{Trivialization of the co-Tangent Bundle}

 We denote by $\mathfrak{so}(3)$ the tangent space at identity
 of the smooth manifold $SO(3)$. Its dual space (the vector space of all
 linear functions on $\mathfrak{so}(3)$) will be denoted by
 ${\mathfrak{so}(3)}^{\ast}$. It is a classical fact that
 the cotangent bundle of $SO(3)$ 
 is \emph{globally} diffeomorphic to the product $SO(3)
 \times {\mathfrak{so}(3)}^{\ast}$. That is, with each pair $(g,P)$
 in $SO(3)  \times {\mathfrak{so}(3)}^{\ast}$, we may associate 
 a linear form $p$ on $T_gSO(3)$ defined by
  $p(v)=P \left ( dR_g^{-1}(v) \right )$.
Using exactly the same 
idea,
 the tangent bundle of $SO(3)$ is diffeomorphic to $SO(3) \times
 \mathfrak{so}(3)$ by identification of $dR_g^{-1}(T_gSO(3))$ with
 $\mathfrak{so}(3)$. Using this formalism,
 the canonical symplectic form on $T^{\ast}SO(3)$ has a nice 
expression (see \cite{jurd-book}), and we may rewrite the PMP in a 
compact  form.

\subsubsection{Hamiltonian of the System}

For the problem ${\cal P}$, the PMP-Hamiltonian $\HHH$ (see \r{HAMHAMHAM}) can be thought as a map
$H_v:{{\frak{so}}}(3)^{\ast}\to \R$ that for each $\lambda_0 \in \R$, and each control $v=(v_1,v_2)\in\R^2$
is defined by
\bqn
\label{EQ-Hamiltonian}
   H_v(P)=P \left [\left ( \begin{array}{ccc}
             0& -v_1 & 0 \\
            v_1 & 0 & -v_2 \\
             0 & v_2 & 0
       \end{array}
      \right )  \right ]+ \lambda_0 (v_1^2+ \frac{1}{\al^2} v_2^2).
\eqn
Using this notation,
the Pontryagin Maximum Principle writes as follows. Notice that, as proved in \cite{q2}, it is enough
to require the transversality condition at the source (and at identity).  \\\\
\noi{\bf Corollary (Pontryagin Maximum Principle for Right Invariant
Problems)}
{\it  Let
$(v(.),g(.))\in L^{\infty}([0, T],\R^2) \times Lip([0, T],SO(3))$ be a 
control-trajectory pair of the system (\ref{EQ-dynsystem}).
If $(v(.),g(.))$ is a solution to the problem ${\cal P}$
then there exists a
constant $\lambda_0 \leq 0$ and a Lipschitzian curve $P(.) \in Lip([0,
T],{\mathfrak{so}(3)}^{\ast})$ such that the pair
$(P(.),\lambda_0)$ never vanishes and for almost every $t$ in $[0,T]$ we have 
\bqn 
\label{EQ-PMP} \left \{
\begin{array}{lcr}
\frac{dg(t)}{dt}&=&dH_{v(t)}(P(t))g(t), \\
\frac{dP(t)}{dt}&=&-ad^{\ast}_{dH_{v(t)}(P(t))}P(t). \end{array}
\right. 
\eqn 
Moreover 
\bqn 
&&H_{v(t)}(P(t))=\max_{\bar v\in
\R^2}  H_{\bar v}(P(t))=const \label{EQ-Max}\\ 
&&P(t)(T_{id}g(t){\So}g^{-1}(t)) =0 \label{EQ-Transversalite}
\textit{ (Transversality condition).}
\eqn
}

\medskip\noi
In the statement of the PMP, $ad^{\ast}_{dH_{v(t)}(P(t))}P(t)$ is
defined by $ad^{\ast}_{dH_{v(t)}(P(t))}P(t)(v)=
P(t)\left ( [dH_{v(t)}(P(t)),v] \right ).$
In this case, since there are no abnormal extremals, (see Proposition 
\ref{p-no-ab}) we
can normalize $\lambda_0=-1/2$. An easy computation leads to
\bqn 
v_1(.)=\phi_1(.)&\mbox{~~and~~}& v_2(.)=\al~ \phi_2(.). \nn
\eqn
where 
\bqn
\phi_1(t)=P(t) \left [\left ( \begin{array}{ccc}
             0& -1 & 0 \\
             1 & 0 & 0 \\
             0 & 0& 0
\ea\right)\right] &\mbox{~~and~~}& \phi_2(t)=\al 
             P(t) \left [\left ( \begin{array}{ccc}
             0& 0 & 0 \\
             0 & 0 & -1 \\
             0 & 1& 0
\ea\right)\right].\nn
\eqn
Notice that $\phi_1(.)$ and $\phi_2(.)$ are defined as in Definition \ref{d-sw-f}. 

Since the group $SO(3)$ is
semi-simple and compact, then the canonical Killing form
$Kill(x,y)=tr(xy)$ is negative definite on the Lie algebra
$\mathfrak{so}(3)$. Hence $-\frac{1}{2}Kill(\cdot,\cdot)$ provides an
isomorphism between $\mathfrak{so}(3)$
 and its dual space ${\mathfrak{so}(3)}^{\ast}$. Using this identification,
  to each linear form $P:{\mathfrak{so}(3)} \rightarrow \R$ in
  ${\mathfrak{so}(3)}^{\ast}$, we can associate a skew symmetric
   matrix $M_P$ of $\mathfrak{so}(3)$ in the following way: by 
definition of $M_P$, for each 
matrix $N$ in $\mathfrak{so}(3)$, we have 
$-\frac{1}{2}Kill(M_P,N)=P(N)$. Equation (\ref{EQ-PMP})
for $P$ can be rewritten in the famous Lax-Poincar\'{e} form (see
\cite{jurd-MCT} for a proof and further discussions) 
\bqn
\label{EQ-LaxPoin} \frac{dM_P(t)}{dt}=[dH_v(P(t)),M_P(t)], 
\eqn
where we have identified ${\mathfrak{so}(3)}^{\ast
\ast}$ with $\mathfrak{so}(3)$.

\subsubsection{Expressions of the Controls}
\llabel{s-ex-co}
Let us define $m_1(.),m_2(.),m_3(.)$ three Lipschitzian functions
from $[0, T]$ to $\R$ by 
\bqn \label{EQ-def-mk} M_P(t)=:
\left(
\begin{array}{ccc} 0 & -m_1(t) & -m_3(t) \\ m_1(t) & 0 & -m_2(t) \\ m_3(t) &
m_2(t) & 0 \end{array} \right), 
\eqn 
for every time $t$ in
$[0,T]$. From the maximality condition (\ref{EQ-Max}), we deduce
that for almost every $t$ in $[0, T]$, we have $m_1(t)=v_1(t)$ and
$m_2(t)=\frac{1}{\al^2} v_2(t)$. From the transversality condition
(\ref{EQ-Transversalite}), we deduce $v_2(0)=0$. Equation
(\ref{EQ-LaxPoin}) writes
\bqn \label{EQ-syst-mk} \left \{ \begin{array}{rcc}
\frac{dv_1(t)}{dt} & = & -m_3(t)v_2(t) \\
\frac{dv_2(t)}{dt} & = & \al^2~ m_3(t)v_1(t) \\
\frac{dm_3(t)}{dt} & = &-\frac{1-\al^2}{\al^2}~ v_1(t)v_2(t). \end{array}
\right. 
\eqn
\brem 
In the case in which $\al=1$ (see \cite{q2}), the system
(\ref{EQ-syst-mk}) admits the solution $v_1(t)=\cos(Kt)$,
$v_2(t)=\sin(Kt)$, $m_3=K$, where $K$ is a real constant. If $\al
\neq 1$ and $m_3(0)\neq0$, the function $m_3(.)$ is no more  constant. 
\erem
In the case in which $\alpha \neq 1$, it is easy to check that the
two functions $K_1(.)$ and $K_2(.)$ defined by 
\bqn
\label{EQ-integral1}
K_1: t\in [0, T] \mapsto  \frac{1}{2}\left(v_1(t)^2 +\frac{1}{\al^2} 
v_2(t)^2\right),~~~K_2: t\in [0,T] \mapsto 
\frac{1}{2}\left(v_1(t)^2-\frac{\al^2}{1-\al^2} m_3(t)^2\right),
\eqn
have zero derivative and hence are constant on the time interval
$[0, T]$. Notice that $K_1(.)$ is just the maximized Hamiltonian 
\r{EQ-Max}.  In the following we normalize 
$K_1=\frac{1}{2}$, that corresponds to parametrize the extremals by arclength. In this way 
our problem is equivalent to a minimum time problem with controls constrained in an ellipse 
(cf. Section \ref{s-intro}).

As already said in Section 4, we restrict our study to trajectories
that remain in the first positive octant $S^+$. For such trajectories,
we want to prove that $m_3(0) \geq 0$.
Since $\psi_2(.)\geq 0$, we must have $v_1(0)\geq 0$. We already know
 (for transversality reasons) that $|v_1(0)|=1$. Hence $v_1(0)=1$.
 Proceed now by contradiction and assume $m_3(0)<0$. Equations (\ref{EQ-syst-mk})
 insure that $v_2(.)$ is strictly negative for (strictly positive) small
 times. This implies that $\psi_3(.)<0$ for small times, which 
contradicts the fact that $\psi(.)$ takes values in $S^+$. 
 As a corollary, 
 we may state that $v_2(.)$ is positive for small times.  
  Using now the classical theory
of elliptic functions (see \cite{Whittaker} chapter XXII for
example), we can express $v_1(.)$, $v_2(.)$ and $m_3(.)$ as
elliptic functions of order 2. Indeed, writing, for all $t$ 
small enough 
\bqn 
v_2(t)=\al \sqrt{1-v_1(t)^2},~~\mbox{ and }~~ 
m_3(t)=\sqrt{m_3(0)^2-\frac{1-\al^2}{\al^2}(1-v_1(t)^2)}, 
\eqn 
we can express $v_1(.)$ as the solution of 
\bqn \label{EQ-elliptic}
\frac{dv_1(t)}{dt}=-\al \sqrt{1-v_1(t)^2}
\sqrt{m_3(0)^2-\frac{1-\al^2}{\al^2}(1-v_1(t)^2)}, 
\eqn 
with
\bqn \label{EQ-initial-condition} v_1(0)=1. 
\eqn 
The solution of (\ref{EQ-elliptic}) with initial condition
(\ref{EQ-initial-condition}) is known for small times (see 
\cite{Whittaker}, chapter XXII). Hence we have the following:
\bp 
For every extremal trajectory, 
starting from $(1,0,0)^T$ 
downstairs (or from $\cal S$ upstairs), there exists $\eps>0$ such that,  
for $t\in[0,\eps]$, it corresponds to controls a.e. 
given by:
 \begin{itemize}
 \item If $m_3(0)=0$, then
         \begin{eqnarray}
	  v_1(t)&=&1, \label{EQ-sol-v1-cas0}\\
	  v_2(t)&=&0,\\
	  m_3(t)&=&0. \label{EQ-sol-m3-cas0}
	  \end{eqnarray}
 \item If $\al <1$ and $m_3(0)^2 < \frac{1-\al^2}{\al^2}$, then
         \begin{eqnarray}
         \label{EQ-sol-v1-cas1} \label{EQ-sol-u1-cas1}
         v_1(t)&=&\mbox{\emph{dn}}
         \left ( \sqrt{1-\al^2}~ t;k \right ), \\
         \label{EQ-sol-v2-cas1}
         v_2(t)&=&\al ~k \mbox{ \emph{sn}} \left (
         {\sqrt{1-\al^2}}~ t;k \right ),\\
	 m_3(t)&=& m_3(0)~{\emph{cn}} \left( \sqrt{1-\al^2}~ t;k \right),
         \label{EQ-sol-m3-cas1}
	 \end{eqnarray} 
         where
         we have defined the modulus of the elliptic functions $dn$, 
         $sn$ and $cn$ to be equal to $k=\frac{\al 
~m_3(0)}{\sqrt{1-\al^2}}.$
 \item If $\al \leq 1$ and $m_3(0)^2 = \frac{1-\al^2}{\al^2}$, then	
         \begin{eqnarray}
         \label{EQ-sol-v1-cas2}
         v_1(t)&= &\frac{1}{\cosh(\sqrt{1-\al^2}~t)},\\
            \label{EQ-sol-v2-cas2}
         v_2(t)&= &\al \tanh(\sqrt{1-\al^2}~t),\\
	 m_3(t)&=& 
\frac{m_3(0)}{\cosh(\sqrt{1-\al^2}~t)}
\label{EQ-sol-m3-cas2}.
         \end{eqnarray}
 \item If $\al \leq 1$ and $m_3(0)^2 > \frac{1-\al^2}{\al^2}$, then
         \begin{eqnarray}
         \label{EQ-sol-v1-cas3} \label{EQ-sol-u1-cas3}
         v_1(t)&=&\mbox{\emph{cn}}
         \left ( \al ~m_3(0) ~t;k \right ),\\
        \label{EQ-sol-v2-cas3}
         v_2(t)&=& \al \mbox{ \emph{sn}} \left (
         \al ~m_3(0) ~t;k \right ),\\
	 m_3(t)&=& m_3(0)~\mbox{\emph{dn}} \left ( \al~ m_3(0)t;k \right 
)\label{EQ-sol-m3-cas3},
	  \end{eqnarray}
	 where
         we have defined the modulus of the elliptic functions $cn$, $sn$ 
and $dn$  to be equal to
        \bqn 
        k=\frac{\sqrt{1-\al^2}}{\al~ m_3(0)}. 
        \eqnl{k1}
 \item If $\al >1$, then
        \begin{eqnarray}
         \label{EQ-sol-v1-cas4} \label{EQ-sol-u1-cas4}
         v_1(t)&=&\mbox{\emph{cd}}
         \left ( \frac{\sqrt{\al^2-1}}{k}~t;k \right ),\\
         \label{EQ-sol-v2-cas4}
         v_2(t)&=& \al \sqrt{1-k^2}~\mbox{\emph{sd}} \left (
         \frac{\sqrt{\al^2-1}}{k}
          ~t;k \right ), \\
	  m_3(t)&=& m_3(0) ~\mbox{\emph{nd}}\left(   
          \frac{\sqrt{\al^2-1}}{k}~t;k   \right), \label{EQ-sol-m3-cas4} 
	  \end{eqnarray}
         where
         we have defined the modulus of the elliptic functions $cd$, $sd$ 
         and $nd$ to be equal to
         \bqn 
         k=\sqrt{\frac{\al^2-1}{\al^2 m_3(0)^2 + \al^2 -1}}.
         \label{EQ-sol-k-cas4}\label{k2} 
         \eqn
\end{itemize}
\llabel{p-cont}
\ep
It happens that the expressions for $v_1(.)$, $v_2(.)$ and $m_3(.)$ 
given in  Proposition \ref{p-cont} make sense for
every time in $\R^+$. Plugging them into  (\ref{EQ-syst-mk}),
one checks that they are in fact solutions for every time in $\R^+$. 
Therefore we have:
\bp
The result of Proposition \ref{p-cont} holds for every $t\in\R^+$.
\llabel{p-cont2}
\ep
\brem
Notice that, in Propositions 
\ref{p-cont} and   \ref{p-cont2}, we do not 
require 
that the trajectory reaches the target.  
To find the optimal  trajectory reaching the target (it happens to be unique),
we first prove that all
these extremals are in fact optimal, and then we determine numerically the 
value of $m_3(0)$ corresponding to it. 
In this way, we also provide the complete optimal synthesis.
\erem

It is easy to check that the map $(t,m_3(0),\al) \mapsto (v_1(t),v_2(t))$
 is continuous. It is also possible to prove that this map is actually analytic.
Consequently, the map $(t,m_3(0),\al) \mapsto \psi(t)$ is itself analytic.
Equations (\ref{EQ-sol-v1-cas0}) to  (\ref{EQ-sol-k-cas4})  give
the form of the optimal controls  in the sense that for every optimal
trajectory $\psi(.)$ starting from point $(1,0,0)^T$, there exists
one $m_3(0)$ such
 that $\psi(.)$ corresponds to controls $v_1(.)$ and $v_2(.)$ as given
 by equations   (\ref{EQ-sol-v1-cas0}) to (\ref{EQ-sol-k-cas4}).
What remains to do is to prove that all these trajectories  are in fact  
optimal in $S^+$.
This is the content of the next section. First we need some preliminary 
results collected in the following 
proposition that is illustrated  in Figure \ref{f-jacobi-total-2}.


\bp\llabel{p-period}
If $(\al,m_3(0))$ satisfies $\al\leq1$ and $m_3(0)^2=\frac{1-\al^2}{\al^2}$, then $v_1(.)$ is a decreasing
function with limit equal to 0 at infinity. If $(\al,m_3(0))$ satisfies $\al>1$ 
or $m_3(0)^2\neq\frac{1-\al^2}{\al^2}$, then $v_1(.)$ is a periodic function  
with period  
${\Theta}_{\al,m_3(0)}$ and it is decreasing until 
$\frac{{\Theta}_{\al,m_3(0)}}{2}$. 
Let us  extend the definition of ${\Theta}_{\al,m_3(0)}$ to couples 
$(\al,m_3(0))$ such
that $\al\leq1$ and $m_3(0)^2=\frac{1-\al^2}{\al^2}$ by setting ${\Theta}_{\al,m_3(0)}=+\infty$.

If $t$, $m_3(0)$ and $\tilde m_3(0)$ are such that $m_3(0)<\tilde m_3(0)$ and
$t\leq \frac{1}{2}\min\{{\Theta}_{\al,m_3(0)},{\Theta}_{\al,\tilde m_3(0)}\}$ then their corresponding
controls satisfy $v_1(t)>\tilde v_1(t)$.
\ep
\proof
The proof of the first claims is a direct consequence of formulas 
\r{EQ-sol-v1-cas0}, \r{EQ-sol-v1-cas1}, \r{EQ-sol-v1-cas2}, \r{EQ-sol-v1-cas3}
and \r{EQ-sol-v1-cas4} for control $v_1(.)$.
Let us prove the last claim. For $t\in[0,{\Theta}_{\al,m_3(0)}/2]$, 
$\dot v_1(t)$ is given by equation \r{EQ-elliptic}. Since the right hand 
side of that equation is strictly decreasing with respect to  $m_3(0)$ for 
$m_3(0)>0$, one concludes, using standard comparison arguments in ODEs. 
\quadp

\begin{figure}
\begin{center}
\input{f-jacobi-total-2.pstex_t}
\caption{Decreasing of $v_1(t)$ as function of $m_3(0)$ and $t$}
\llabel{f-jacobi-total-2}
\end{center}
\end{figure}
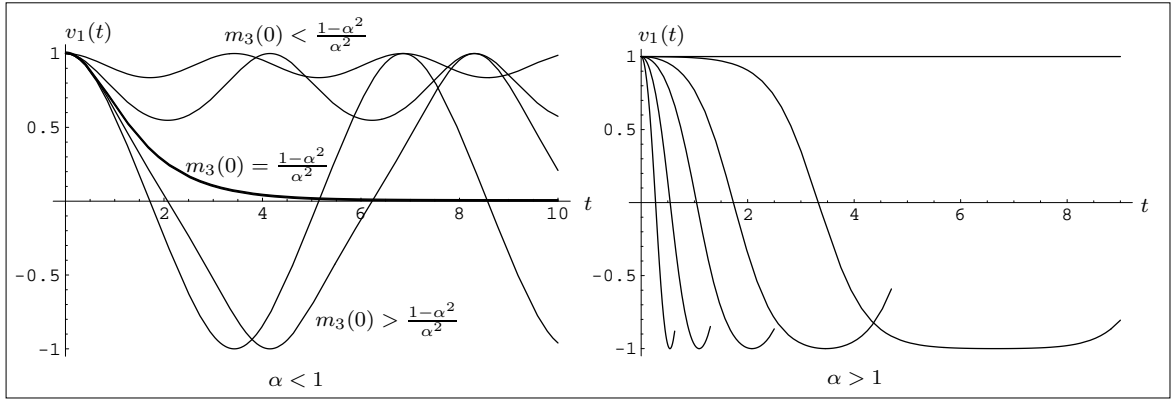




 \subsection{Optimal Synthesis} \llabel{SEC-OptimalSynthesisSubRiem}
  
In this section, we prove that all the extremals given by
Propositions \ref{p-cont} and \ref{p-cont2} are actually optimal as long 
as 
they stay in  $S^+$.

\brem
Recall that in $S^+$, our minimization problem is a singular-Riemannian 
problem.  In $S^+\setminus\da$, the singular-Riemannian structure is in 
fact a Riemannian one. In spherical coordinates
$(\theta,\varphi)$, its curvature is 
$$
c(\theta,\varphi)=   \frac{\left( -4 (1+\al^2) + 
      \left( -1 + \al^2 \right) \,
       \left( 2\,\cos (2\,\varphi) + 
         \cos (2\,\left( \varphi - \theta \right) ) + 
         \cos (2\,\left( \varphi + \theta \right) ) \right) 
      \right) \,{\cos(\theta)}^2}{4\,
    \al^2\sin(\theta)^4}<0.
$$
If we were exactly in the context of Riemannian geometry, since the 
topology of $S^+$ is trivial, this would be enough to conclude  the 
non existence of cut-locus or conjugate locus, i.e., that every extremal 
trajectory is 
in fact optimal. 
Here, the structure is singular along $\da$ and a correct proof 
of optimality via curvature
requires some care and leads to tedious details. For this reason 
in what follows we prefer to prove optimality without using the concept of 
curvature. 
\erem       
The first step is 
to prove that there is
no cut-locus. 

\subsubsection{Non Existence of Cut-Locus in ${\cal S}^+$} 

Recall (see Section \ref{s-prel}) that in $S^+$, the functions 
$f_1(.)$ and
$f_2(.)$ are  defined by: 
$ f_1(\psi)= -\frac{\psi_1}{\psi_2},$ $f_2(\psi)=\alpha 
\frac{\psi_3}{\psi_2}.$
With the same techniques used in Section \ref{s-mt}, (cf. Proposition \ref{cap2:s1}) one can easily prove
that, since $f_1(.)<0$ (resp. $f_2(.)>0$ ) in $Int(S^+)$, then $v_1(.)$ (resp. $v_2(.)$) can change 
of sign only once from positive to negative (resp. from negative to positive).
Since for small values of $t$, $v_2(t)$ is
positive, 
we have that $v_2(.)$ remains positive (\emph{resp.} strictly positive) 
as long as the corresponding 
trajectory belongs to $S^+$ (\emph{resp.} the interior of $S^+$).
In particular, this implies that for every  extremal trajectory, 
$\psi_3(.)$ is an increasing function of the time. Moreover since $m_3(.)$ 
is positive (see equations (\ref{EQ-sol-m3-cas0}),
 (\ref{EQ-sol-m3-cas1}), (\ref{EQ-sol-m3-cas2}), (\ref{EQ-sol-m3-cas3}) and
 (\ref{EQ-sol-m3-cas4})), the system (\ref{EQ-syst-mk}) insures that $v_1(.)$
 is decreasing as long as the extremal trajectory stays in $S^+$. Hence we 
have:
\bp\llabel{p-sign}

Let $(\psi(.),v(.))$, be a trajectory-control pair 
given by Propositions 
\ref{p-cont}, \ref{p-cont2}, starting from $(1,0,0)^T$. Then $\psi(.)$ 
leaves $S^+$ in finite time. More precisely setting   
$\psi(.)=(\psi_1(.),\psi_2(.),\psi_3(.))$ there exists $\bar t\in\R^+$ 
such that $\psi_1(\bar t)=0$ or $\psi_2(\bar t)=0$ and $\psi(t)\in S^+$ in 
$[0,\bar t]$. Moreover in  $[0,\bar t]$ we have:\\
$\bullet$ $v_2(t)\geq0$.\\
$\bullet$ $\psi_3(.)$ is an increasing function.\\
$\bullet$ $v_1(.)$ is decreasing.\\
Finally every extremal trajectory $\psi(.)$ cannot self-intersect in 
$S^+$.
 In other words, for every  $s<t<\bar t$ 
we have  $\psi(s) \neq \psi(t)$.
\ep
Now, let us investigate the structure of extremal curves for small time.
Using the Hamiltonian equations \r{EQ-syst-mk},
 it is easy to compute the 3-jet of the trajectory $\psi(.)$ corresponding 
to a given $m_3(0)$.
 One gets that $\psi(0),$  $\dot\psi(0)$ and  $\ddot\psi(0)$ do not depend 
on 
$m_3(0)$, but one has $\stackrel{\ldots}{\psi_1}(0)=0$, 
$\stackrel{\ldots}{\psi_2}(0)= -\al^2m_3(0)^2-1 $ 
and $\stackrel{\ldots}{\psi_3}(0)= 2 \al^2 ~m_3(0).$ 
It follows 
that, two curves, corresponding 
to two different values of $m_3(0)$,
do not intersect for strictly positive (small enough) time.

\bp \label{PRO-CutLocusRiem-Int}
There is no cut locus in $S^+$. 
More precisely for every positive $t$ such that $\psi(t)$ and
$\tilde{\psi}(t)$ belong to $S^+$ and such that 
$\psi(.)$ and $\tilde{\psi}(.) $ are both optimal up to time $t$, we have 
$\psi(t) \neq \tilde{\psi}(t)$.
\ep

\proof  
First notice that, since by hypotheses the two curves are optimal and  we 
are considering only curves starting in $S^+$, then  their supports are 
entirely contained in $S^+$. 

Consider two optimal curves $\psi(.)$ and 
$\tilde{\psi}(.)$, corresponding respectively to $m_3(0)$, 
$v_1(.)$, $v_2(.)$  and 
$\tilde{m}_3(0)$, $\tilde{v}_1(.)$, 
$\tilde{v}_2(.)$. Since the jets of $\psi(.)$ and $\tilde\psi(.)$ are 
different at time zero, it follows that there exists $\eps>0$ such that 
$Supp(\psi|_{]0,\eps[})\cap Supp(\tilde\psi|_{]0,\eps[})=\emptyset$.

If  the supports of the two curves do not intersect in $S^+$, then there 
is nothing to prove. Assume now that the supports of the two curves 
intersect in $S^+$. Let $t$ be the first positive time such that 
$Supp(\psi|_{]0,t]})\cap Supp(\tilde\psi|_{]0,t]})\neq\emptyset$. 
Since the curves are parametrized by 
arclength, and they are optimal, it follows that $\psi(t)=\tilde\psi(t)$.

\medskip
Let us first consider the case where $\psi(t) \in S^+
\setminus \Delta_A^{-1}(0)$. We may assume that $m_3(0)<\tilde{m_3}(0)$.
Since $\tilde{\psi}(.)$ is optimal, it does not self intersect and reaches 
the boundary of $S^+$ in finite time. Hence, the support of 
$\tilde{\psi}(.)$ delimits two simply connected compact domains in 
$S^+$. In the following we call $S_R^+$, the region  containing points of 
$\da$ arbitrarily close to $(1,0,0)^T$, and $S_L^+$ the other one.
Let us prove that $\dot{\psi}(t)$ cannot be 
collinear to $\dot{\tilde{\psi}}(t)$.
Indeed, on one hand,
if these vectors are collinear with the same versus at $t$, then
$v_1(t)=\tilde{v}_1(t)$. 
But, thanks to Proposition \ref{p-sign}, 
since the two curves are optimal and hence extremal in $S^+$,
their controls $v_1(.)$ and $\tilde{v}_1(.)$ are decreasing until $t$. 
Hence, thanks to Proposition \ref{p-period}, $t$ is less than their half 
periods and $v_1(t)=\tilde{v}_1(t)$ implies $m_3(0)=\tilde{m}_3(0)$. 
On the other hand, if these vectors are collinear with opposite versus, then
$v_2(t)=-\tilde{v}_2(t)=0$ which implies that $\psi(t)$ and 
${\tilde{\psi}}(t)$ belong to $\Delta_{B_2}^{-1}(0)$. But one can 
easily 
check that the only optimal trajectory with points in $\dbd$, for $t>0$, 
corresponds to $m_3(0)=0$. Let us now prove that ${\psi(t)}$ cannot be 
transverse to 
$\tilde{\psi}(t)$. Indeed, since $m_3(0)<\tilde{m_3}(0)$ then, for small 
time $s$, $\psi(s)$ belongs 
to $S^+_L$. Moreover, since $t$ is the first time at which $\psi(.)$ and 
$\tilde{\psi}(.)$ intersect, then $Supp(\psi)|_{]0,t[}\subset S^+_L$. 
Hence 
the only way for 
$\dot{\psi}(t)$ to be 
transverse to $\dot{\tilde{\psi}}(t)$ is to point toward $S^+_R$.
But this
implies that $v_1(t)<\tilde{v}_1(t)$, which is in contradiction with 
Proposition \ref{p-period}.
Hence we got the contradiction for the existence of cut-points in $S^+\setminus\da$.

If $\psi(t)\in\da$, the previous argument can 
be adapted in the following way. Computing the 3-jets of $\psi(.)$ and 
$\tilde{\psi}(.)$ at time $t$, one can prove that 
$Supp(\psi|_{]0,t[})\subset S^+_R$. Moreover, computing the 3-jets of 
$\psi(.)$  and
$\tilde{\psi}(.)$  at time zero, one can prove that 
$Supp(\psi|_{]0,t[})\subset S^+_L$. It follows the contradiction.
\quadp

\subsubsection{Optimality of Extremals}

Proposition \ref{PRO-CutLocusRiem-Int} allows to prove:

\bp \llabel{PRO-Synt-Opt-SubRiem}
Each extremal trajectory issued from $(1,0,0)^T$ is optimal as long as it 
stays in  $S^+$. As a consequence, by each point of $S^+$ passes only one 
extremal trajectory such that its support is entirely included in $S^+$. 
\ep

\proof In $S^+$, we define the set  $\cal D$ by ${\cal D}=
 \Delta_{B_1}^{-1}(0) \cup \Delta_{A}^{-1}(0)\setminus\{ (1,0,0)^T\}$.
 On $\cal D$, we define
the complete order $<<$ by:\\
$\bullet$ $(0,\psi_2,\psi_3) << (0, \tilde{\psi_2}, \tilde{\psi_3})$ iff
 $\psi_2 > \tilde{\psi_2}$,\\
$\bullet$    $(0,\psi_2,\psi_3) << (\tilde{\psi_1}, 0, \tilde{\psi_3})$ if 
$\psi_2 >0$,\\
$\bullet$  $(\psi_1,0,\psi_3) << (\tilde{\psi_1}, 0, \tilde{\psi_3})$ iff 
$\psi_1 < \tilde{\psi_1}$.\\
From Proposition \ref{PRO-CutLocusRiem-Int}, we know that for every point 
$x$ in $\cal D$,
there is only one optimal trajectory issued from $(1,0,0)^T$ that reaches $x$. 
We define the function $\Phi(.)$ which, to each point in $\cal D$, 
 associates the value $m_3(0)$ of the corresponding optimal trajectory.
Our goal is to show that  $\Phi(.)$ is an homeomorphism between ${\cal D}$ 
and $\R^+.$

Optimal trajectories cannot intersect, hence $\Phi(.)$ is an increasing 
function. Now, we claim that $\Phi(.)$ is actually continuous.
Indeed since it is increasing, then, for every $x\in {\cal D}$, 
$\Phi(.)$ 
has an upper limit $\ell(x)$ on
the left and a lower limit $L(x)$ on the right at $x$. 
If $\Phi(.)$ were not continuous at $x$, then $\ell(x)<L(x)$. But both 
extremals corresponding to 
$m_3(0)=\ell(x)$ and $m_3(0)=L(x)$ reach $x$ and are optimal as limit of equicontinuous optimal curves.
This is in contradiction with Proposition \ref{PRO-CutLocusRiem-Int}. 
Hence $\Phi$ is continuous.

As a consequence, since $\Phi$ is a continuous increasing function, it 
is one-to-one from $\cal D$
to a certain interval  $I$ of $\R^+$. 
Since $\psi_3(.)$ is a strictly increasing function in the interior of 
$S^+$, the 
trajectory associated to $m_3(0)=0$, $v_1(.)=1$, $v_2(.)=0$ is the only  
extremal reaching  the point $(0,1,0)^T$ and hence it is optimal. This 
proves that $0 \in I$.
Assume now that $I$ is bounded. The knowledge of the jets of the extremals corresponding
to $m_3(0)$ in $I$ shows that there exists an $\epsilon>0$ such that every 
extremal corresponding to
$m_3(0)\in I$ stays in $S^+$ until $\epsilon$. But thanks to the continuity of the distance function to
$(1,0,0)^T$, there are points in $\da\setminus\{(1,0,0)^T\}$ whose 
subriemannian distance to 
$(1,0,0)^T$ is less than $\epsilon$.
Hence these points are not reached by curves with $m_3(0)\in I$ 
which is a contradiction.  Hence $I$ is not bounded.

Hence, to each value of $m_3(0)$, it corresponds one trajectory $\psi(.)$ in $S^+$ 
and one point $x$ in $\cal D$ such that $\psi(.)$ is the only extremal reaching 
$x$. We conclude that $\psi(.)$ is optimal 
in $S^+$. \quadp



\subsubsection{Minimum Time of Transfer}
\llabel{s-trans-time}

In this section we provide explicit expressions of the minimum time needed to reach 
the target from the source, as function of the nonisotropy factor $\al$ 
and of the
value of $m_3(0)$, that will be determined numerically in the next section.

Recall that Proposition \ref{PRO-Synt-Opt-SubRiem} insures that there is only one extremal, reaching 
the target $(0,0,1)^T$ that remains in the positive octant. Since, at 
$(0,0,1)^T$, we have $F_1=0$, then $v_1(T)=0.$
 
Actually, thanks to Proposition \ref{p-period}, $T$ is
the \emph{first} root of $v_1(.)$. Moreover, it can be seen from equations 
\r{EQ-sol-v1-cas0}, 
\r{EQ-sol-v1-cas1} and \r{EQ-sol-v1-cas2}, that,  if $m_3(0)=0$, or if  
$\al\leq 1$ and $m_3(0)^2\leq \frac{1-\al^2}{\al^2}$, then $v_1(.)$ has no 
root.
Hence, we have the following expressions for the minimum time of transfer as function of $\al$ and $m_3(0)$.
In the case $\al=1$, $m_3(0)$ and the minimum time of transfer have been computed in \cite{q2}.
Set ${\bf K}(k):=\int_0^{\frac{\pi}{2}} \frac{ds}{\sqrt{1-k^2 
\sin^2(s)}}$, i.e., the complete elliptic integral of the second kind. We 
have:
\bqn
&&\mbox{{\bf Case $\al<1$.}}~~ T=\frac{{\bf K}(k)}{\al~ m_3(0)},~~\mbox{ 
where $k$ is given by \r{k1}.}\label{Tm1}\\
&&\mbox{\bf Case $\al=1$.}~~m_3(0)=\frac{1}{\sqrt 3}\mbox{ and}:~~ 
T=\sqrt3\frac{\pi}{2}.\\
&&\mbox{\bf Case $\al>1$.}~~ T=\frac{{\bf K}(k)}{\sqrt{\al^3m_3(0)^2+\al^2-1}},
~~\mbox{ where $k$ is given by  \r{k2}.}\llabel{Tp1}
\eqn

\subsubsection{Upper Bound and Numerical Estimation of $m_3(0)$}
\llabel{s-numerical-shit}

In this section, we provide a numerical estimation of the initial condition $m_3(0)$ for
the optimal trajectory reaching the target $(0,0,1)^T$. The method 
consists in using a
dichotomy algorithm in the following way.

Thanks to Proposition \ref{PRO-Synt-Opt-SubRiem}, optimal 
trajectories associated to $\tilde{m}_3(0)>m_3(0)$ 
leave $S^+$ at a point
of the set $\{ \psi_2=0 \}$ and optimal trajectories associated to $\tilde{m}_3(0)<m_3(0)$
leaves $S^+$ at a point of the set $\{ \psi_1=0 \}$. 
Hence, starting from an arbitrary value of $m_3(0)$, at each step,
the next value will be chosen smaller if the corresponding optimal trajectory leaves
$S^+$ at a point of the set $\{ \psi_2=0 \}$ and larger in the other case.
This algorithm is of course much more efficient if we provide an a priori 
upper bound 
of $m_3(0)$. 

Recall that the problem of minimizing energy in fixed time is equivalent 
to minimize time with controls constrained in the set
$V_\al:=\{(v_1,v_2)\in\R^2:~~v_1^2+v_2^2/\al^2\leq1\}$. 
Moreover recall that for the isotropic case $\al=1$, the time to reach 
the target $(0,0,1)^T$ is $\frac{\sqrt{3} \pi}{2}$. More in general,
for an isotropic case with  controls constrained 
in the dilated set 
$W^{\mbox{{\small 
isotr}}}_{\al'}:=\{(v_1,v_2)\in\R^2:~~v_1^2+v_2^2\leq\al'^2\}$, $\al'>0$,
the time to reach
the target is $\frac{\sqrt{3} \pi}{2\al'}$. In the following we use the 
fact that if $V_\al\subset W^{\mbox{{\small isotr}}}_{\al'}$, then  
the time to reach the target $T$ is bigger than $\frac{\sqrt{3} 
\pi}{2\al'}$. We have:

\medskip
\noindent {\bf Case $\al<1$.}
From equation \r{Tm1}, we deduce $T \leq 
\frac{\pi}{2~\al~m_3(0)} \frac{1}{\sqrt{1-k^2}},$ where 
$k$ is given by formula \r{k1}. Moreover since in this case $V_\al\subset 
W^{\mbox{{\small isotr}}}_1$ we get $m_3(0)\leq  \frac{1}{\al\sqrt{3}}
\frac{1}{\sqrt{1-k^2}}\leq\sqrt{\frac{4}{3 \al^2}-1}.$

\medskip

\noindent {\bf Case $\al=1$.}
$m_3(0)=\frac{1}{\sqrt3}$

\medskip

\noindent {\bf Case $\al>1$.}
From equation \r{Tp1}, we deduce $\sqrt{\al^2m_3(0)^2+\al^2-1}~ T \leq 
\frac{\pi}{2}  
\frac{1}{\sqrt{1-k^2}},$
where $k$
is given by formula \r{k2}. Since in this case,  
 $V_\al\subset
W^{\mbox{{\small isotr}}}_\al$ we get $T \geq \frac{\sqrt{3} \pi}{2 \al}$ 
and hence $m_3(0) \leq \frac{1}{\sqrt{3}}.$
Figure \ref{f-numerika} 
depicts the time needed to reach the target and the 
corresponding value of $m_3(0)$ as functions of the nonisotropy factor 
$\al$.

\ppotR{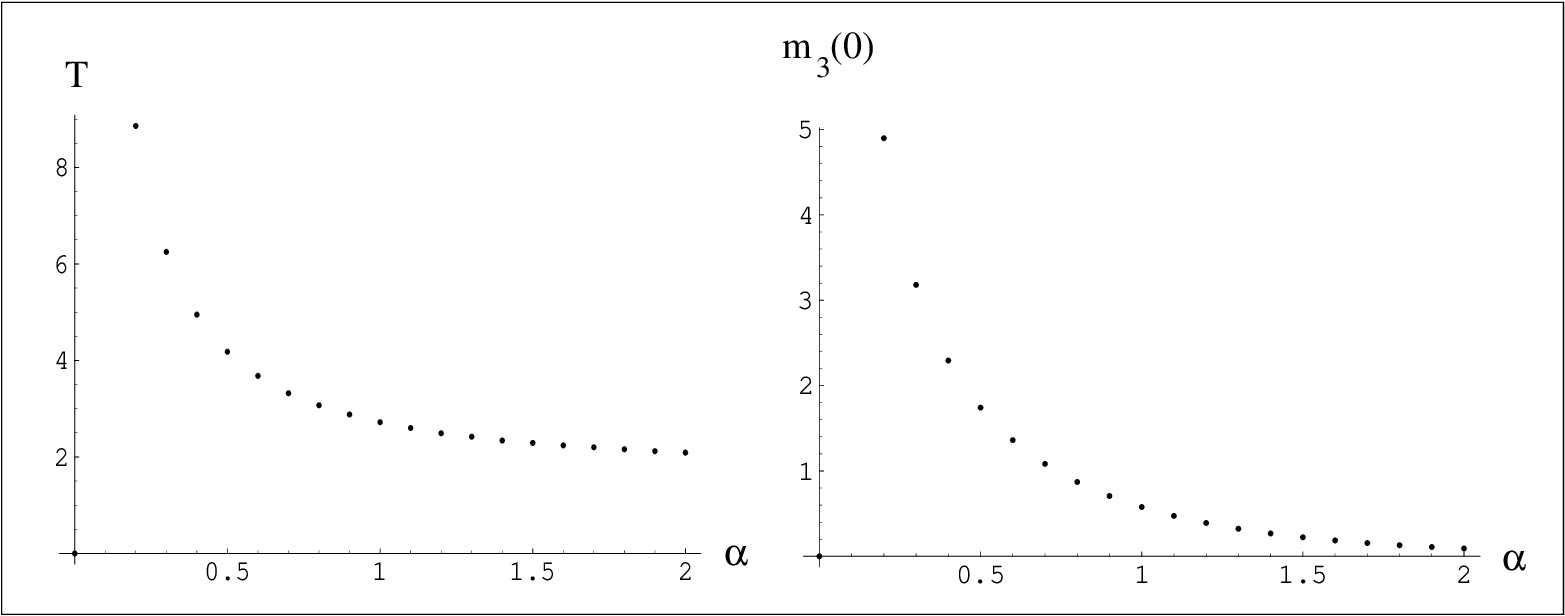}{
Time needed to reach the target and 
corresponding value of $m_3(0)$ as function of the nonisotropy factor
$\al$.}{12}

\begin{figure}
\begin{center}
~\includegraphics[width=15truecm]{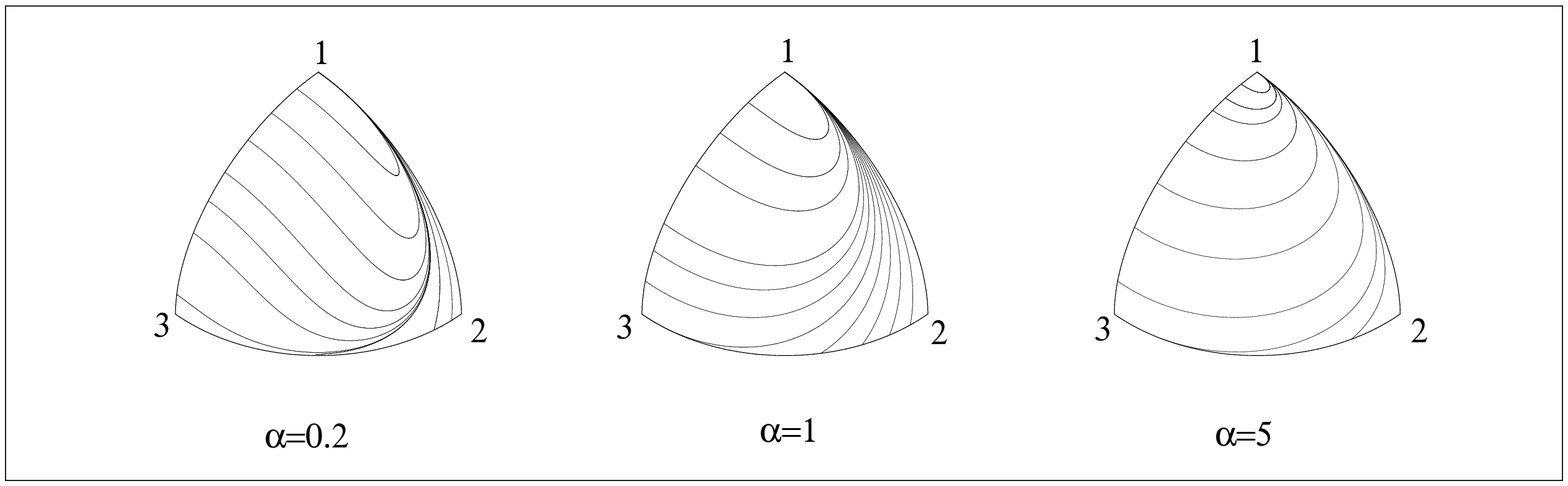}
\caption{Images of the optimal synthesis for the minimum energy 
problem for $\al=0.2$, $\al=1$ and $\al=5$}
\llabel{f-nonnepossopiu}
\end{center}
\end{figure}

\end{document}

%% file: f-leq.pstex_t
\begin{picture}(0,0)%
\includegraphics{f-leq.pstex}%
\end{picture}%
\setlength{\unitlength}{1579sp}%
\begingroup\makeatletter\ifx\SetFigFont\undefined%
\gdef\SetFigFont#1#2#3#4#5{%
  \reset@font\fontsize{#1}{#2pt}%
  \fontfamily{#3}\fontseries{#4}\fontshape{#5}%
  \selectfont}%
\fi\endgroup%
\begin{picture}(9654,5619)(4,-6028)
\put(5131,-856){\makebox(0,0)[lb]{\smash{{\SetFigFont{8}{9.6}{\rmdefault}{\mddefault}{\updefault}{\color[rgb]{0,0,0}$F_1$}%
}}}}
\put(3991,-1546){\makebox(0,0)[lb]{\smash{{\SetFigFont{8}{9.6}{\rmdefault}{\mddefault}{\updefault}{\color[rgb]{0,0,0}$F_1$}%
}}}}
\put(2266,-3031){\makebox(0,0)[lb]{\smash{{\SetFigFont{8}{9.6}{\rmdefault}{\mddefault}{\updefault}{\color[rgb]{0,0,0}$[F_1,F_2]$}%
}}}}
\put(3541,-3541){\makebox(0,0)[lb]{\smash{{\SetFigFont{8}{9.6}{\rmdefault}{\mddefault}{\updefault}{\color[rgb]{0,0,0}$[F_1,F_2]$}%
}}}}
\put(6121,-4291){\makebox(0,0)[lb]{\smash{{\SetFigFont{8}{9.6}{\rmdefault}{\mddefault}{\updefault}{\color[rgb]{0,0,0}$F_2$}%
}}}}
\put(5971,-5641){\makebox(0,0)[lb]{\smash{{\SetFigFont{8}{9.6}{\rmdefault}{\mddefault}{\updefault}{\color[rgb]{0,0,0}$F_2$}%
}}}}
\put(1951,-1141){\makebox(0,0)[lb]{\smash{{\SetFigFont{8}{9.6}{\rmdefault}{\mddefault}{\updefault}{\color[rgb]{0,0,0}$\da$}%
}}}}
\put(6631,-991){\makebox(0,0)[lb]{\smash{{\SetFigFont{8}{9.6}{\rmdefault}{\mddefault}{\updefault}{\color[rgb]{0,0,0}$\dbd$}%
}}}}
\put(7156,-2611){\makebox(0,0)[lb]{\smash{{\SetFigFont{8}{9.6}{\rmdefault}{\mddefault}{\updefault}{\color[rgb]{0,0,0}$f_1>0,~f_2<0$}%
}}}}
\put(1006,-5701){\makebox(0,0)[lb]{\smash{{\SetFigFont{8}{9.6}{\rmdefault}{\mddefault}{\updefault}{\color[rgb]{0,0,0}$\dbu$}%
}}}}
\end{picture}%

%% file: f-synt.pstex_t
\begin{picture}(0,0)%
\includegraphics{f-synt.pstex}%
\end{picture}%
\setlength{\unitlength}{1895sp}%
\begingroup\makeatletter\ifx\SetFigFont\undefined%
\gdef\SetFigFont#1#2#3#4#5{%
  \reset@font\fontsize{#1}{#2pt}%
  \fontfamily{#3}\fontseries{#4}\fontshape{#5}%
  \selectfont}%
\fi\endgroup%
\begin{picture}(16824,6114)(124,-5953)
\put(14506,-3649){\makebox(0,0)[lb]{\smash{{\SetFigFont{8}{9.6}{\rmdefault}{\mddefault}{\updefault}{\color[rgb]{0,0,0}$(1,1)$}%
}}}}
\put(12121,-2644){\makebox(0,0)[lb]{\smash{{\SetFigFont{8}{9.6}{\rmdefault}{\mddefault}{\updefault}{\color[rgb]{0,0,0}$(-1,1)$}%
}}}}
\put(10696,-2116){\makebox(0,0)[lb]{\smash{{\SetFigFont{8}{9.6}{\rmdefault}{\mddefault}{\updefault}{\color[rgb]{0,0,0}$(1,0)$}%
}}}}
\put(9841,-3856){\makebox(0,0)[lb]{\smash{{\SetFigFont{8}{9.6}{\rmdefault}{\mddefault}{\updefault}{\color[rgb]{0,0,0}$(1,1)$}%
}}}}
\put(7576,-2086){\makebox(0,0)[lb]{\smash{{\SetFigFont{8}{9.6}{\rmdefault}{\mddefault}{\updefault}{\color[rgb]{0,0,0}$(-1,1)$}%
}}}}
\put(5086,-2192){\makebox(0,0)[lb]{\smash{{\SetFigFont{8}{9.6}{\rmdefault}{\mddefault}{\updefault}{\color[rgb]{0,0,0}$(1,0)$}%
}}}}
\put(1951,-2942){\makebox(0,0)[lb]{\smash{{\SetFigFont{8}{9.6}{\rmdefault}{\mddefault}{\updefault}{\color[rgb]{0,0,0}$(-1,1)$}%
}}}}
\put(4456,-4052){\makebox(0,0)[lb]{\smash{{\SetFigFont{8}{9.6}{\rmdefault}{\mddefault}{\updefault}{\color[rgb]{0,0,0}$(1,1)$}%
}}}}
\put(736,-5297){\makebox(0,0)[lb]{\smash{{\SetFigFont{8}{9.6}{\rmdefault}{\mddefault}{\updefault}{\color[rgb]{0,0,0}$(0,1)$}%
}}}}
\put(11116,-255){\makebox(0,0)[lb]{\smash{{\SetFigFont{8}{9.6}{\rmdefault}{\mddefault}{\updefault}{\color[rgb]{0,0,0}$\g^{++}$}%
}}}}
\put(15691,-1219){\makebox(0,0)[lb]{\smash{{\SetFigFont{8}{9.6}{\rmdefault}{\mddefault}{\updefault}{\color[rgb]{0,0,0}$(1,0)$}%
}}}}
\put(5731,-409){\makebox(0,0)[lb]{\smash{{\SetFigFont{8}{9.6}{\rmdefault}{\mddefault}{\updefault}{\color[rgb]{0,0,0}$\g^{++}$}%
}}}}
\put(2416,-5791){\makebox(0,0)[lb]{\smash{{\SetFigFont{8}{9.6}{\rmdefault}{\mddefault}{\updefault}{\color[rgb]{0,0,0}$\al<1$}%
}}}}
\put(8296,-5760){\makebox(0,0)[lb]{\smash{{\SetFigFont{8}{9.6}{\rmdefault}{\mddefault}{\updefault}{\color[rgb]{0,0,0}$\al=1$}%
}}}}
\put(13756,-5749){\makebox(0,0)[lb]{\smash{{\SetFigFont{8}{9.6}{\rmdefault}{\mddefault}{\updefault}{\color[rgb]{0,0,0}$\al>1$}%
}}}}
\put(10351,-5761){\makebox(0,0)[lb]{\smash{{\SetFigFont{8}{9.6}{\rmdefault}{\mddefault}{\updefault}{\color[rgb]{0,0,0}switching points}%
}}}}
\put(10351,-5536){\makebox(0,0)[lb]{\smash{{\SetFigFont{8}{9.6}{\rmdefault}{\mddefault}{\updefault}{\color[rgb]{0,0,0}curve made of}%
}}}}
\end{picture}%

%% file: f-jacobi-total-2.pstex_t
\begin{picture}(0,0)%
\includegraphics{f-jacobi-total-2.pstex}%
\end{picture}%
\setlength{\unitlength}{1776sp}%
\begingroup\makeatletter\ifx\SetFigFont\undefined%
\gdef\SetFigFont#1#2#3#4#5{%
  \reset@font\fontsize{#1}{#2pt}%
  \fontfamily{#3}\fontseries{#4}\fontshape{#5}%
  \selectfont}%
\fi\endgroup%
\begin{picture}(16269,5544)(1429,-5428)
\put(2266,-391){\makebox(0,0)[lb]{\smash{{\SetFigFont{8}{9.6}{\rmdefault}{\mddefault}{\updefault}{\color[rgb]{0,0,0}$v_1(t)$}%
}}}}
\put(5101,-5251){\makebox(0,0)[lb]{\smash{{\SetFigFont{8}{9.6}{\rmdefault}{\mddefault}{\updefault}{\color[rgb]{0,0,0}$\al<1$}%
}}}}
\put(3946,-2251){\makebox(0,0)[lb]{\smash{{\SetFigFont{8}{9.6}{\rmdefault}{\mddefault}{\updefault}{\color[rgb]{0,0,0}$m_3(0)=\frac{1-\al^2}{\al^2}$}%
}}}}
\put(9511,-2791){\makebox(0,0)[lb]{\smash{{\SetFigFont{8}{9.6}{\rmdefault}{\mddefault}{\updefault}{\color[rgb]{0,0,0}$t$}%
}}}}
\put(5761,-4411){\makebox(0,0)[lb]{\smash{{\SetFigFont{8}{9.6}{\rmdefault}{\mddefault}{\updefault}{\color[rgb]{0,0,0}$m_3(0)>\frac{1-\al^2}{\al^2}$}%
}}}}
\put(4486,-436){\makebox(0,0)[lb]{\smash{{\SetFigFont{8}{9.6}{\rmdefault}{\mddefault}{\updefault}{\color[rgb]{0,0,0}$m_3(0)<\frac{1-\al^2}{\al^2}$}%
}}}}
\put(17281,-2821){\makebox(0,0)[lb]{\smash{{\SetFigFont{8}{9.6}{\rmdefault}{\mddefault}{\updefault}{\color[rgb]{0,0,0}$t$}%
}}}}
\put(12901,-5221){\makebox(0,0)[lb]{\smash{{\SetFigFont{8}{9.6}{\rmdefault}{\mddefault}{\updefault}{\color[rgb]{0,0,0}$\al>1$}%
}}}}
\put(10261,-421){\makebox(0,0)[lb]{\smash{{\SetFigFont{8}{9.6}{\rmdefault}{\mddefault}{\updefault}{\color[rgb]{0,0,0}$v_1(t)$}%
}}}}
\end{picture}%

%% file: after-review.bbl
\begin{thebibliography}{99}      

\bbibitem{agra-book} A.A. Agrachev, Yu.L. Sachkov,  {\it Control Theory 
from the Geometric Viewpoint}, Encyclopedia of Mathematical Sciences, 
Vol.87, Springer, 2004.


\bibitem{mario}  A.A. Agrachev,  M. Sigalotti,  {\it On the
Local
Structure of Optimal Trajectories in $\R^3$}, {SIAM J. Control
Optimization}, Vol. 42, No. 2,  pp.
513--531, 2003.






\bbibitem{G-a}  R. El Assoudi, J.P. Gauthier, I.A.K.  Kupka, {\it On 
subsemigroups of semisimple Lie groups}, Ann. Inst. H. Poincar\e' Anal. 
Non 
Lin\'eaire 13  No. 1, pp. 117--133, 1996. 


\bbibitem{bellaiche} A. Bellaiche, {\it The tangent space in 
sub-Riemannian
geometry}, Sub-Riemannian geometry,  Progr.
Math., Vol. 144, Birkh\"auser, Basel, pp. 1--78, 1996. 


\bbibitem{bts} K. Bergmann, H. Theuer and B.W. Shore, {\it Coerent
population transfer among quantum states of atomes and molecules}, Rev.
Mod. Phys. Vol. 70, pp. 1003--1025, 1998.

\bbibitem{boltianski} V,G, Boltyanskii, {\it Sufficient Conditions for 
Optimality and the Justification of the Dynamics Programming Principle}, 
SIAM J. Control and Optimization, Vol. 4, pp. 326--361, 1996.


\bbibitem{bonnard-book} B. Bonnard, M. Chyba, {\it The Role of Singular 
Trajectories in Control Theory}, Springer, SMAI, Vol. 40, 2003.



\bbibitem{automaton}
U. Boscain and B. Piccoli, {\it On automaton recognizability
of abnormal extremals}, SIAM J. Control and Optimization, Vol. 40 No.5
pp. 1333--1357, 2002.



\bbibitem{libro} U. Boscain, B, Piccoli, {\it Optimal Synthesis  for
Control Systems on 2-D Manifolds,} Springer, SMAI, Vol. 43, 2004.


\bibitem{q1} U. Boscain, G. Charlot, J.-P. Gauthier,
St\'ephane Gu\'erin and Hans--Rudolf Jauslin, {\it Optimal Control in
laser-induced population transfer for two- and three-level quantum
systems}, Journal of Mathematical Physics,  Vol. 43, pp. 2107--2132, 2002.          



\bbibitem{q2} U. Boscain, T. Chambrion and J.-P. Gauthier, 
{\it On the K+P problem for a three-level quantum system: 
Optimality implies resonance}, Journal of Dynamical and 
Control Systems, Vol. 8,  pp. 547--572, 2002.

\bbibitem{q3} U. Boscain, G. Charlot, {\it Resonance of Minimizers for
$n$-level Quantum Systems with an Arbitrary Cost}, 
 ESAIM: Control, Optimisation and Calculus of Variations (COCV), Vol. 10, 
pp. 593--614, 2004. 


\bbibitem{sevilla} U. Boscain, T. Chambrion and J.-P. Gauthier, 
{\it Optimal Control on a $n$-level Quantum System}, 
 Lagrangian
and Hamiltonian methods for nonlinear control, pp.  129--134, 2003.  

\bbibitem{dubin}  U. Boscain, Y. Chitour, {\it Time Optimal Synthesis for 
Left--Invariant Control Systems on SO(3)}, to appear in SIAM  J. Control 
and Optimization, 2005.


\bbibitem{quattro} A. Bressan and B. Piccoli, {\it A
Generic Classification of Time Optimal Planar Stabilizing Feedbacks},
SIAM J. Control and Optimization, Vol.  36
No.1,  pp. 12-32, 1998.


\bbibitem{brun1} P. Brunovsky,  {\it Existence of Regular Syntheses for
General Problems,}    J. Diff. Equations  Vol. 38,
pp. 317-343, 1980.


\bbibitem{cesari} L. Cesari, {\it Optimization-theory and applications:
problems with
ordinary differential equations}, New York,  Springer-Verlag, 1983.


\bbibitem{rabitz} M. A. Dahleh, A. P. Peirce and H. Rabitz, {\it Optimal
control of quantum-mechanical systems: Existence, numerical approximation,
and applications}, Phys. Rev. A,  Vol. {37}, pp. 4950--4964, 1988.


\bbibitem{daless} D. D'Alessandro and M. Dahleh, {\it Optimal control of
two-level
quantum systems},
  IEEE Transactions on Automatic Control, Vol. 46, No. 6, 
pp. 866--876. 2001.


\bbibitem{car1}  U. Gaubatz, P. Rudecki, M. Becker, S. Schiemann, M. 
Kulz, and K. Bergmann, {\it Population switching
                     between vibrational levels in molecular beams,} 
Chem. Phys. Lett. Vol. 149, pp. 463--468, 1988.



\bbibitem{G-gb} J.P. Gauthier and G. Bornard. {\it Controlabilite des 
sytemes bilineaires}, SIAM J. Control and Optimization, Vol. 20, 
pp. 377--384, 1982.




\bbibitem{jurd-book} V. Jurdjevic, {\it Geometric Control Theory},
Cambridge University Press, 1997.      

\bbibitem{jurd-MCT} V. Jurdjevic, {\it Optimal Control, Geometry and
Mechanics}, Mathematical Control Theory, J. Bailleu, J.C. Willems (ed.),
pp. 227--267, Springer, 1999.


\bbibitem{G-jk} V. Jurdjevic and I.K. Kupka, {\it Control Systems
on Semisimple Lie Groups and Their Homogeneous Spaces}, {\sl Ann. Inst.
Fourier}, Vol. 31, pp. 151--179, 1981.


\bbibitem{brokko} N. Khaneja, R. Brockett and S. J. Glaser {\it Time
optimal control in spin systems}, Phys. Rev. A
Vol. {63}, 032308, 11 pp., 2001.



\bbibitem{car2} C. Liedenbaum, S. Stolte, J. Reuss, {\it Inversion 
Produced and Reversed by Adiabatic Passage}, Phys Rep. Vol. 178, pp. 
1--24, 1989.

\bbibitem{montgomery} R. Montgomery, {\it A Tour of Subriemannian
Geometry}, American Mathematical Society, Mathematical Surveys and
Monographs, 2002. 


\bbibitem{due} B. Piccoli, {\it Classifications 
of Generic Singularities
for the Planar
Time-Optimal Synthesis}, {\sl SIAM J. Control and Optimization,} Vol. 34
No. 6, pp. 1914--1946, 1996.


\bbibitem{piccoli-sussmann} B. Piccoli and H.J. Sussmann, {\it Regular
Synthesis and Sufficiency Conditions for Optimality},  SIAM J.
Control and Optimization, Vol. 39, No. 2, pp. 359--410, 2000.


\bbibitem{pontlibro} L.S. Pontryagin, V. Boltianski, R. Gamkrelidze and E.
Mitchtchenko, {\it The Mathematical Theory of Optimal
Processes},  John Wiley and Sons, Inc, 1961.  




\bbibitem{rama} V. Ramakrishna, H. Rabitz, M.V. Salapaka, M. Dahleh and 
A. Peirce. {\it Controllability of Molecular Systems.}, Phys. Rev. A, 
Vol. 62, pp. 960--966, 1995.

\bbibitem{yuri} Y. Sachkov, {\it Controllability of Invariant Systems on 
Lie Groups and 
Homogeneous Spaces,} J. Math. Sci., Vol. 100, No. 4, pp. 2355--2427, 2000.


\bbibitem{sch1} H. Schattler, {\it On the Local Structure of Time Optimal
Bang-Bang Trajectories in $\R^3$}, {SIAM J. Control Optimization.}
  Vol. 26, No. 1,  pp.
186--204, 1988. 


\bbibitem{shorebook} Shore, Bruce W., {\it The theory of coherent atomic 
excitation.} New York, NY, Wiley, 1990.



\bbibitem{sus1} H.J. Sussmann, {\it The structure of time-optimal
trajectories for single-input systems in the plane: the general real
analytic case},  SIAM J. Control Optimization, Vol. 25, No. 4,
pp. 868--904, 1987.



\bbibitem{sus2} H.J. Sussmann, {\it Regular synthesis for time optimal
control of single--input real--analytic systems in the plane}, {\sl
SIAM J.  Control and Optimization,} Vol. 25, pp. 1145--1162, 1987.


\bibitem{Whittaker} E.T. Whittaker and G.N. Watson, \textit{A
course of Modern Analysis}, Cambridge University Press, 1927.




\end{thebibliography}
